\documentclass[footinbib,a4paper,aps,prb,reprint,twocolumn,preprintnumbers,amsmath,amssymb,10pt, superscriptaddress]{revtex4-2}
\usepackage{verbatim}
\usepackage{color}
\usepackage{tikz}
\usepackage{xr}
\usepackage[colorlinks=true,citecolor=blue,linkcolor=blue,urlcolor=blue]{hyperref}
\usepackage{braket}
\usepackage[normalem]{ulem}
\usepackage{upgreek}
\usepackage[percent]{overpic}
\usepackage{graphicx,xcolor}
\usepackage{dcolumn}
\usepackage{bm}
\usepackage[shortlabels]{enumitem}
\usepackage{bbold}

\newcommand{\Pa}{\mathcal{P}_{\alpha}}

\renewcommand{\rm}[1]{\textrm{#1}}

\definecolor{forestgreen}{rgb}{0.1, 0.6, 0.2}

\makeatletter
\renewcommand{\fnum@figure}{FIG. \thefigure}
\makeatother

\begin{document}

\title{Diagrammatic method for many-body non-Markovian dynamics: memory effects and entanglement transitions}
\author{Giuliano Chiriac\`{o}}
\email{giuliano.chiriaco@dfa.unict.it}
\affiliation{Dipartimento di Fisica e Astronomia ``Ettore Majorana'', Universit\`{a}
di Catania, 95123 Catania, Italy}
\affiliation{The Abdus Salam International Centre for Theoretical Physics (ICTP), Strada Costiera 11, 34151 Trieste, Italy}
\affiliation{SISSA — International School of Advanced Studies, via Bonomea 265, 34136 Trieste, Italy}
\author{Mikheil Tsitsishvili}
\affiliation{The Abdus Salam International Centre for Theoretical Physics (ICTP), Strada Costiera 11, 34151 Trieste, Italy}
\affiliation{SISSA — International School of Advanced Studies, via Bonomea 265, 34136 Trieste, Italy}
\author{Dario Poletti}
\affiliation{Science, Mathematics and Technology Cluster, Singapore University of Technology and Design, 8 Somapah Road, 487372 Singapore}
\affiliation{Engineering Product Development Pillar, Singapore University of Technology and Design, 8 Somapah Road, 487372 Singapore}
\affiliation{The Abdus Salam International Centre for Theoretical Physics (ICTP), Strada Costiera 11, 34151 Trieste, Italy}
\affiliation{Centre for Quantum Technologies, National University of Singapore 117543, Singapore}
\author{Rosario Fazio}
\affiliation{The Abdus Salam International Centre for Theoretical Physics (ICTP), Strada Costiera 11, 34151 Trieste, Italy}
\affiliation{Dipartimento di Fisica, Universit\`{a} di Napoli “Federico II”, Monte S. Angelo, I-80126 Napoli, Italy}
\author{Marcello Dalmonte}
\affiliation{The Abdus Salam International Centre for Theoretical Physics (ICTP), Strada Costiera 11, 34151 Trieste, Italy}
\affiliation{SISSA — International School of Advanced Studies, via Bonomea 265, 34136 Trieste, Italy}

\date{\today}

\begin{abstract}
We study the quantum dynamics of a many-body system subject to coherent evolution and coupled to a non-Markovian bath. We propose a technique to unravel the non-Markovian dynamics in terms of quantum jumps, a connection that was so far only understood for single-body systems. We develop a systematic method to calculate the probability of a quantum trajectory, and formulate it in a diagrammatic structure. We find that non-Markovianity renormalizes the probability of realizing a quantum trajectory, and that memory effects can be interpreted as a perturbation on top of the Markovian dynamics. We show that the diagrammatic structure is akin to that of a Dyson equation, and that the probability of the trajectories can be calculated analytically. We then apply our results to study the measurement-induced entanglement transition in random unitary circuits. We find that non-Markovianity does not significantly shift the transition, but stabilizes the volume law phase of the entanglement by shielding it from transient strong dissipation.
\end{abstract}

\maketitle

\section{Introduction}

Quantum systems in the real world are subject to their own coherent evolution as well as interactions with the environment. The interplay between these two gives rise to complex and rich physics that has great relevance in the context of quantum technologies, and has consequently been extensively studied in recent years. It is the case of many solid state, cold atoms or trapped ions systems, where external interactions can drive a transition \cite{Diehl08,verstraete2009quantum,Sieberer16,Lee13,Jin16,Maghrebi2016,FossFeig2017,Fink17,Fitzpatrick17,Flaschner18,Syassen08,Marino16,Rota19,Young20,Marcuzzi14,seetharam2021correlation}, such as in dissipative phase transitions tuned by the dissipation strength, or induce new relaxation regimes \cite{Poletti12,Poletti13,Sieberer13,SciollaKollath2015,Schiro16,Chiriaco18,Chiriaco2020,Chiriaco20b,Sun20}. This interplay is important also in the context of quantum information where, for example, systems can decouple from the incoherent action of the environment and form dissipative free subspace with important error-preventing properties \cite{Lidar1998:DFS,Lidar1999:DFS,Plenio1997:DFS,Bacon2000:DFS,Beige2000:DFS,Kwiat2000:DFSexp,DeFilippo2000:DFS,Shabani2005:DFS,Lidar2003:bookDFS}. 

Dissipative phase transitions occur at the level of the average state -- i.e. manifest themselves in the properties of the density matrix of the system -- but new phases may emerge also at the level of single quantum trajectories, as highlighted by a series of recent works \cite{Dhar16,Nahum2017, Li18, Li2019,Li21, Zhou2019:nahumRUC, Skinner2019, Bao2020, Jian2020:RUC,Gullans2020, Gullans2020a,Gopalakrishnan20,Turkeshi20,Turkeshi2021a,Ippoliti21,Buchhold2021,Minato21,Block2021,Sharma2022,Chen20,Biella2020,Tang21,Jian21, Muller2021,Lunt21,Alberton21,Nahum21,Sierant2022,Cao19,Szyniszewski20,Lang15,Botzung2021,Coppola2022}. Focusing mainly on systems amenable to be realized in cold atoms or quantum computing platforms, these works have shown that the competition between quantum measurements and coherent dynamics also gives rise to transitions of the entanglement that manifest themselves in specific observables - such as von Neumann entropies, negativities, or two-time correlation functions. These transitions are often referred to as “measurement induced” phase transitions (MIPT). 

All works published so far consider measurements or baths that are Markovian. While many experiments can still be adequately modeled using Markovian baths, this remains an important issue: both natural and engineered baths are most times non-Markovian \cite{Breuer:Petruccione,Rivas2014:NM_measuresReview,Breuer2016:NM_measuresReview,Breuer2009:NM_measures,Wolf2008:NM_measures,Rivas2010:NM_measures,Hou2011:NM_measures,Usha2011:NM_measures,Bylicka2014:NM_measures,Chrucinski2014:NM_measures,Pineda2016:NM_measures,He2017:NM_measures,Xu2022:NM_measures,Ask2022:NM_measures, GuoPoletti2020, Guo2022:NM_measures,Glick2020:NM_QM,Link2022:NMcavity,Flannigan2022:NM_QSDdaley,Chruscinski2022:NM_QJdesign_Piilo,Dann2022:NM_localMasterEq, vanKampen1998, WisemanGambetta2008,LiWiseman2018}, and the backflow of information from the bath into the system is inevitably present in realistic systems and may have dramatic consequences, but its effects are yet to be explored.

In this work, we investigate the consequences of a non-Markovian bath on the dynamics of many-body systems, and, within those, the effects of information backflow on the entanglement transition. We achieve this goal by presenting a new methodology to perform the unraveling of the non-Markovian dynamics of many-body systems in terms of quantum jumps, and showing a specific application of this method to a random unitary circuit featuring an entanglement transition.

The study of non-Markovian systems is broad and challenging~\cite{Breuer2016:NM_measuresReview}. Even at the level of the density matrix, it is not always possible to describe the dynamics through a Lindblad equation. This subject has been extensively studied in the literature, including its many connections to complexity and entanglement \cite{Rivas2010:NM_measures,Milz2018:NM_entanglement,Abiuso2022:NM_measures}, and how to quantify the degree of non-Markovianity etc. \cite{Breuer2009:NM_measures,Wolf2008:NM_measures,Rivas2010:NM_measures, GuoPoletti2020,Guo2022:NM_measures}. In this manuscript we choose to work with a paradigmatic model, in which the dynamics of the system is described by a master equation of the type
\begin{equation}\label{Eq:NMQMasterEquationIntro}
\dot{\rho}(t)=\mathcal{L}_t\rho(t).
\end{equation}
The time dependent Liouvillian $\mathcal{L}_t$ depends on the details of the unitary evolution and on a time dependent dissipation rate, which may be either positive or negative depending on the direction of the information flow. Information flows from the system to the environment and the decay rate is positive for Markovian regions, while it goes from the bath to the system (with an associated negative rate) when the evolution is non-Markovian.

Another difficulty of studying the entanglement of many-body non-Markovian systems is the need to consider the quantum trajectories of its dynamics, and so far no clear and general protocol to unravel such non-Markovian dynamics exists. Unlike in Markovian systems, where the unraveling is performed in a straightforward way using methods such as Monte Carlo wave function (MCWF) \cite{MolmerDalibard1993, DalibardMolmer1992, Daley2014}, or Quantum State Diffusion (QSD)\cite{WisemanMilburn2009}, unraveling recipes have proven to be much harder to implement for non-Markovian systems.

In this work, we tackle these challenges and formulate a description of non-Markovian many-body dynamics in terms of quantum trajectories; this is the backbone of our work and the most challenging task of our analysis.

While in recent years  a protocol implementing the unraveling through quantum jumps has been proposed for single-body systems \cite{Piilo2008,Piilo2009}, still no general approach to many-particle systems exists. Indeed, extending the method of non-Markovian quantum jumps to many-body systems is not trivial, since the interplay between unitary evolution and measurements (which generally compete against each other) makes this task very complicated at a conceptual level, and exponentially complex at a computational level. Just to cite an example, the quantum trajectories of a non-Markovian system are inter-dependent of each other, due to the memory of the bath, and a statistical sampling of the trajectory ensemble similar to the Markovian case is not possible anymore.

In our work, we overcome these technical and conceptual difficulties. We show that the probability of the dynamics realizing a certain quantum trajectory can be calculated analytically when the information backflow restores the information previously lost by the system. The crucial observation we make is that the quantum state of the system (that is, the labeling of each quantum trajectory) only depends on how much and when information was lost by the system without being restored; processes in which information is lost and then flows back into the system do not affect its physical state. Therefore only the unrestored jumps have a physical meaning (and affect the system) and may be detected. Oppositely, the restored jumps do not - in the sense that we can label all trajectories without taking them into account.

Within the above framework, the non-Markovian regions of the evolution renormalize the probability of a trajectory due to the (infinitely) many instances in which information subtracted from the system is later restored through information backflow. Remarkably, such infinite sum can be written in a diagrammatic form, in which it exhibits the exact same formal structure of the Dyson equation for the Greens function of an interacting system, thus providing with an analytic expression for the probability of any trajectory. Beyond being of key practical help in terms of computation (that we exploit in full in the context of random circuits), this unexpected connection between non-Markovian trajectories and Dyson series allows to understand the effect of memory as a “renormalization'' on top of the Markov case. In particular, it establishes a direct correspondence between the unraveling of a non-Markovian Lindblad equation, and the unraveling of a Markovian master equation in terms of trajectories, whose associated ensemble probabilities are determined by the Dyson equation above. We emphasize that, while this feat is generically possible via artificial extensions of the Hilbert space \cite{Imamoglu1994:HilbertExtension,Garraway1996:HilbExt,Breuer1999:HilbExt,Breuer2004:HilbExt}, our mapping is fundamentally distinct, as it identifies a correspondence within the same Hilbert space, thanks to the systematic simplifications enabled by the diagrammatics.

We then apply our formalism to a one dimensional non-Markovian random unitary circuit. The results of our non-Markovian quantum jumps formulation still depend on knowing the quantum state of the system along a certain trajectory, which implies a (practically impossible) full simulation of the system. However, in many settings of random circuits \cite{Jian2020:RUC} the no-measurements probability becomes independent of the quantum state of the system. This allows us to perform further analytical calculations and makes such systems the ideal study case for analyzing the robustness of the entanglement transition. Our formalism enables us to straightforwardly generalize known results \cite{Zhou2019:nahumRUC,Jian2020:RUC} to the non-Markovian case, and perform a mapping to a two-dimensional classical Potts model, where the couplings between spins are inhomogeneous along the time direction. We perform classical Monte Carlo simulations on such a model, finding that the effect of non-Markovianity is to strengthen the volume law phase in the entanglement transition.

The rest of the paper is organized as follows. In Section \ref{Sec:MBnonMarkovQJ} we summarize the quantum jump protocol for single body non-Markovian systems \cite{Piilo2008,Piilo2009} and present a generalization to the many-body case, which is our first main result. In Section \ref{Sec:Diagrams} we calculate the probability of the system dynamics realizing a certain trajectory, and show that the additional contribution due to non-Markovianity can be calculated analytically. This is our second main result, summarized in Eq. \eqref{Eq:PalphaMultiChan}. In Section \ref{Sec:nMEntanglement} we investigate a random unitary circuit subject to non-Markovian measurements. We show that the results of Section \ref{Sec:Diagrams} can be applied to circuits in order to map them to a classical Potts model on which, to study the entanglement transition, we perform Monte Carlo calculations. The Monte Carlo simulations constitute our third main result. Finally, in Section \ref{Sec:Conclusions} we present our conclusions.

\section{Unraveling of many-body non-Markovian dynamics}\label{Sec:MBnonMarkovQJ}

The quantum dynamics of a large class of non-Markovian systems can be described using a time-convonlutionless approximation \cite{Breuer:Petruccione,Piilo2008,Piilo2009,Dann2022:NM_localMasterEq, SmirnePiilo2020, Chruscinski2022:NM_QJdesign_Piilo}. thus, even if a local in time master equation is not the most general description of the dynamics of a non-Markovian system, Eq. \eqref{Eq:NMQMasterEquationIntro} is still an excellent framework to study non-Markovian measurement induced transitions.

For many-body systems with multiple decay channels, the corresponding generalized Gorini–Kossakowski–Sudarshan–Lindblad (GKSL) master equation \cite{Gorini1976, Lindblad1976} reads:
\begin{equation}\label{Eq:NMQMasterEquation}
\dot{\rho}(t)=\frac1i[H,\rho(t)]+\sum_s\Delta_s(t)\left[a_s\rho a^{\dagger}_s-\frac12\{a^{\dagger}_sa_s,\rho\}\right],
\end{equation}
where $H$ is the Hamiltonian of the system, $a_s$ is the jump operator relative to the quantum channel $s$ and $\Delta_s(t)$ is its associated decay rate.

In order to study measurement induced transitions, one has to unravel the dynamics of the system -- i.e. follow the evolution of the state along a single trajectory corresponding to a particular realization of the random quantum jumps. A standard technique is that of the MCWF method \cite{MolmerDalibard1993, DalibardMolmer1992}, where the random quantum jumps are realized through the application of Kraus operators, and the different realizations of jump sequences result in a stochastic ensemble of wave functions corresponding to the quantum states, whose average at any time equals the density matrix of the system. More precisely, after a quantum jump in channel $s$ occurs at time $t$, the system jumps from $|\psi\rangle$ to $|\psi'\rangle$ with a probability $p^{s,+}$:
\begin{gather}\label{Eq:MQJqs_Intro}
|\psi\rangle\rightarrow|\psi'\rangle=\frac{a_s|\psi\rangle}{||a_s|\psi\rangle||};\qquad p^{s,+}\propto\Delta_s(t).
\end{gather}

We see that adapting this recipe to non-Markovian systems presents some problems. An evident issue is that for the times when $\Delta_s(t)<0$ the jump probability would become negative, which has no physical meaning.

\begin{figure}[!t]
    \centering
    \includegraphics[width=0.96\columnwidth]{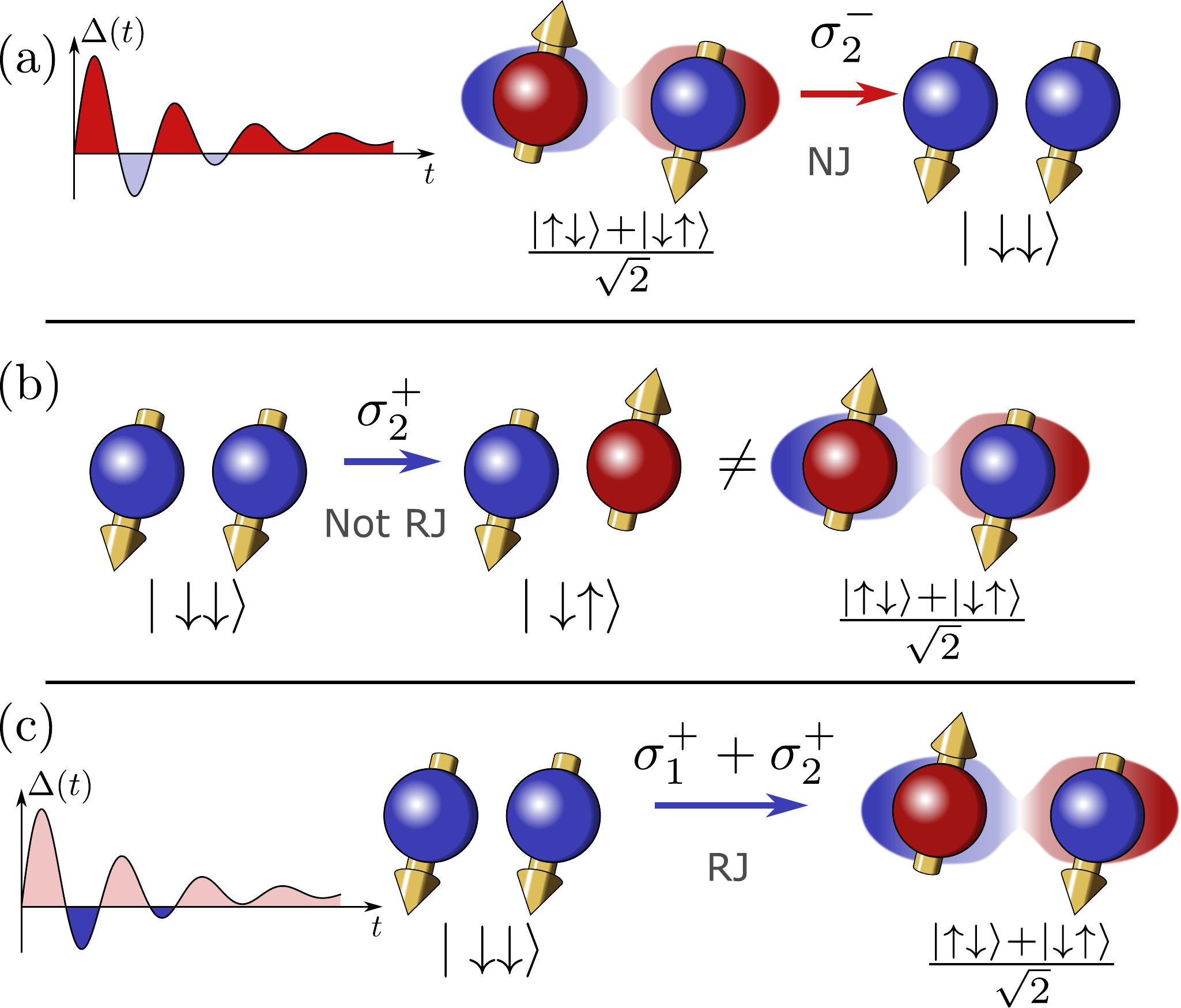}
    \caption{(a) Scheme of an entangled spin pair undergoing a normal jump (NJ) through the jump operator $\sigma_2^-$. (b) Trying to implement the reverse jump (RJ) through $\sigma_2^+$ re-excites the second spin, but does not take the spins into the initial entangled state, and the system stays separable. (c) Instead, in order to restore the original entanglement one has to apply $\sigma_1^++\sigma_2^+$ even though the NJ knew nothing about the presence of spin $1$.}
    \label{fig:NJRJ}
\end{figure}

Another issue is that the back-flow of information from the environment to the system restores not only the population of the excited states, but also coherences, i.e. the off-diagonal elements in the density matrix, as can be seen by solving the master equation \cite{Piilo2008,Piilo2009}. This cannot be implemented through an ``opposite'' jump operator that connects two states in the opposite direction of the corresponding normal jump operator. Figure \ref{fig:NJRJ} shows the simple example of a spin pair that loses its entanglement upon the application of a normal jump operator; applying the inverse operator does not result in retrieving the lost entanglement. This is true even for single body systems: take for example a two level system with $a=\sigma^-$; one may be tempted to use $\sigma^+$ to reverse the effect of the quantum jump, but it can be seen that the application of $\sigma^+$ leads to an increase of the population in the excited state, which still results in a decay of coherences. Restoring quantum coherence is an operation that requires memory of the past evolution of the system, a property that a simple implementation in terms of Kraus operators does not have.

\subsection{Non-Markovian quantum jumps of single body systems}\label{Sec:singlebodyNMQJ}

Before studying the many-body case, we now review in detail what is known about one- (or few-) body systems. This is instructive to highlight the conceptual differences with respect to Markovian dynamics, as well as to identify the major technical challenges that we will address in the many-body case below.

A technique to unravel non-Markovian dynamics has been proposed in Ref.~\cite{Piilo2008,Piilo2009}, in the form of the non-Markovian quantum jumps (NMQJ) method. This prescription allows to describe each interaction with the environment (either Markovian or non-Markovian) in terms of a quantum jump process, and gives back the correct starting master equation \eqref{Eq:NMQMasterEquation} when averaged over the stochastic ensemble of quantum trajectories. 

The main feature of the NMQJ method is the introduction of two different types of quantum jumps in the stochastic evolution of the state: a ``normal jump'' (NJ) occurring during the Markovian regions of the dynamics ($\Delta(t)\geq0$), and a ``reverse'' quantum jump (RJ) acting during the non-Markovian regions ($\Delta(t)<0$). A reverse jump essentially brings the quantum state back to what it was prior to the last Markovian normal jump, effectively cancelling out its effects on the system. More formally, this is described by stating explicitly the probability of performing a jump (either normal or reverse) in the Markovian and non-Markovian regions, and the corresponding initial and final quantum state before and after the jump, similar to Eq. \eqref{Eq:MQJqs_Intro}.

Similarly to the MCWF method, the evolution of the state $\ket{\psi(t)}$ along a trajectory is deterministic, until a random quantum jump occurs. We discretize the evolution of the system, so that the probability that more than one jump occurs within each time interval $\delta t$ is negligible, and assume the jumps to occur instantaneously. The average over the stochastic ensemble gives back the density matrix: $\rho(t)=\sum_{\{\ket{\psi}\}}\frac{N_{\ket{\psi}\bra{\psi}}(t)}{N}\ket{\psi(t)}\bra{\psi(t)}$, where $N_{\ket{\psi}\bra{\psi}}(t)$ is the population of the trajectory $\ket{\psi(t)}$ and $N$ is the total population of the states in the ensemble -- i.e. the ratio $\frac{N_{\ket{\psi}\bra{\psi}}(t)}{N}$ is the stochastic probability of realizing the trajectory $\ket{\psi(t)}$.

Let us now consider a time $t$ for which a particular channel $s$ has a positive rate $\Delta_s(t)>0$. The system may perform a normal jump from a state $\ket{\psi}$ to a state $\ket{\psi'}$ with probability $p^{s,+}$. Explicitly rewriting Eq. \eqref{Eq:MQJqs_Intro}
\begin{gather}\label{Eq:MQJqs} |\psi(t)\rangle\rightarrow|\psi'(t)\rangle=\frac{a_s|\psi(t)\rangle}{||a_s|\psi(t)\rangle||};\\
\label{Eq:MQJprob}p^{s,+}=\Delta_s(t)\delta t\langle \psi(t)|a^{\dagger}_sa_s|\psi(t)\rangle.
\end{gather}

The final state is renormalized, and the probability to perform the jump is given by the decay rate times the probability that the system is in a state eligible to perform the jump, i.e. $\langle \psi(t)|a^{\dagger}_sa_s|\psi(t)\rangle$.

During a non-Markovian region a RJ may occur cancelling out the effect of the last normal jump:
\begin{gather}\label{Eq:NMQJqs}
|\psi'(t)\rangle\leftarrow|\psi(t)\rangle=\frac{a_s|\psi'(t)\rangle}{||a_s|\psi'(t)\rangle||};\\
\label{Eq:NMQJprob} p^{s,-}_{|\psi\rangle\rightarrow|\psi'\rangle}=\frac{N'(t)}{N(t)}|\Delta_s(t)|\delta t\langle \psi'(t)|a^{\dagger}_sa_s|\psi'(t)\rangle.
\end{gather}

The $\leftarrow$ means that the system performs the reverse jump starting from the initial state $|\psi(t)\rangle$, which is the result of applying the NJ operator $a_s$ to the final state $|\psi'(t)\rangle$ after the RJ. This corresponds to effectively erasing the last NJ, and the initial states eligible to reverse jump are the ones that have previously performed (at least) one normal jump. This is also reflected in the expectation value of $a^{\dagger}_sa_s$, which expresses the probability that a certain state is eligible to jump, and that for a reverse jump is calculated on the target state but using the normal jump operators.

\begin{figure}[!t]
    \centering
    \includegraphics[width=\columnwidth]{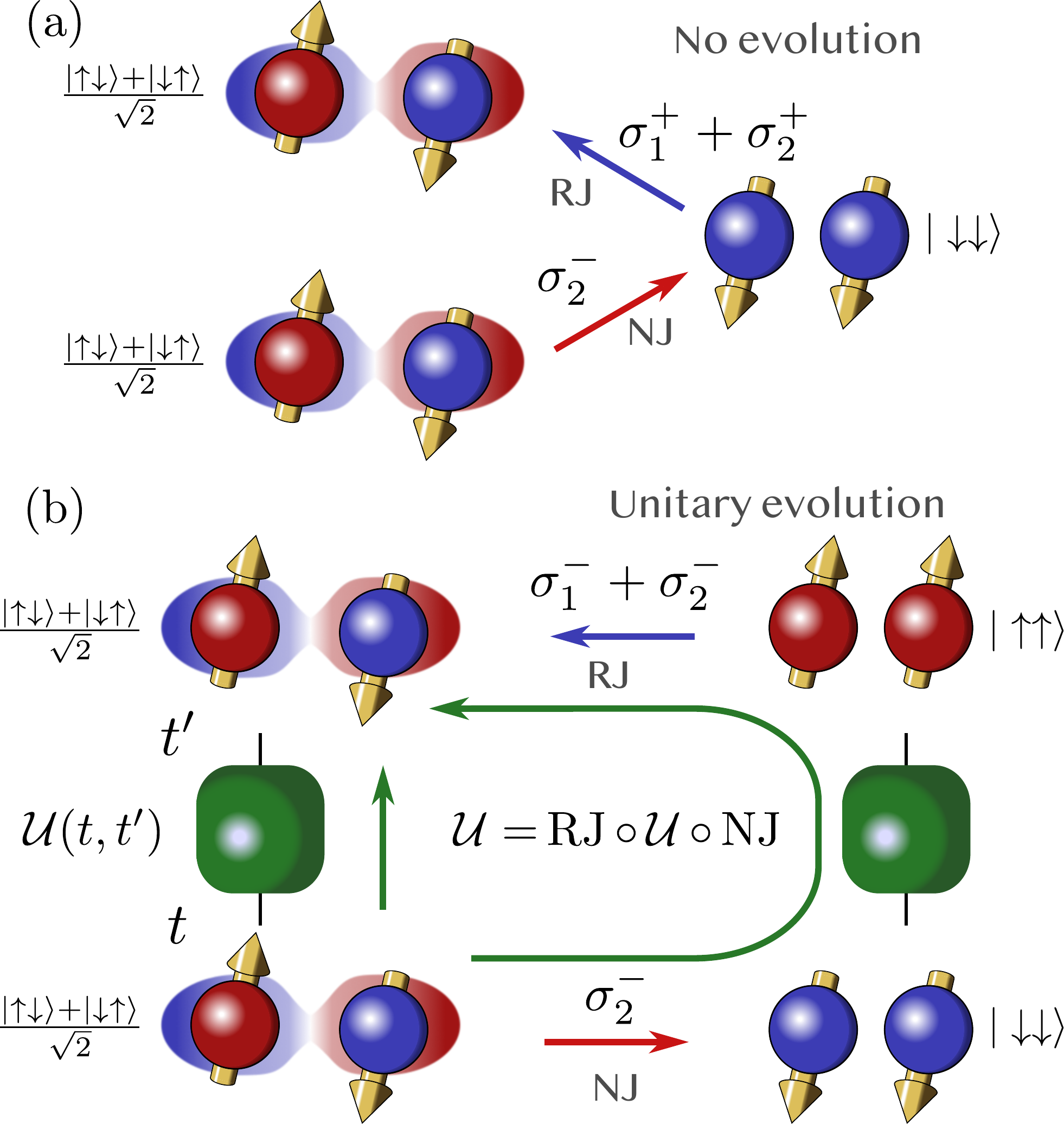}
    \caption{Simple example of the dependence of the reverse jump (RJ) operator on the time at which is performed. A system starts in an entangled Bell pair state and normal jumps (NJ) into the separable state through $\sigma_2^-$. (a) If the RJ occurs immediately, the original state is restored through $\sigma_1^++\sigma_2^+$. (b) If the system evolves from time $t$ to $t'$ through the unitary operator $\mathcal{U}=\sigma_1^x\sigma_2^x$, the separable state is now flipped and the RJ operator that restores the original state is now $\sigma_1^-+\sigma_2^-=\mathcal{U}(\sigma_1^++\sigma_2^+)\mathcal{U}^{\dagger}$.}
    \label{fig:NJRJMB}
\end{figure}

The process in Eq. \eqref{Eq:NMQJqs} cannot be described using a Kraus operator, but is formally obtained by applying the (state dependent) jump operator $\ket{\psi'(t)}\bra{\psi(t)}$. This is a fundamental difference with the MCWF method and a consequence of the memory of the non-Markovian dynamics: the operator corresponding to a RJ depends on the current quantum state and on the target state, see Fig. \ref{fig:NJRJMB}.

Another consequence is the presence of the ratio $N'(t)/N(t)$ in the jump probability: it corresponds to the ratio between the probability of being in the target state and the probability of being in the initial state. This ratio ensures that the evolution averaged over trajectories is described by the master equation Eq. \eqref{Eq:NMQMasterEquation}.

In Refs. \cite{Piilo2008,Piilo2009} the number of inequivalent trajectories -- in the sense that they correspond to different quantum states (we explain it more in detail later on) -- is finite and very small, due to the single body nature of the considered systems. This makes a numerical simulation of the system dynamics viable, since one only has to follow those few trajectories and update the ensemble statistics based on the type of quantum jump performed by the system.

The situation is very different for a many-body system: if the jump operators take the system into states that are not eigenstates of the unitary evolution given by $H$, then the time at which a jump is performed becomes important, resulting in different trajectories. The number of trajectories is then very large, being exponential in the time of the evolution. For example, this occurs when the many-body Hamiltonian contains terms that counteract the action of the jump operators, since at any time the system may or may not decay, and after a decay it may be excited again by the unitary evolution. A very simple example is a two spin-$1/2$ system with jump operators $\sigma_{1/2}^-$ and unitary evolution operator $\sigma_1^x\sigma_2^x$: the steady state for the jump operator has both spins down in the $z$ direction; this is not an eigenstate of the Hamiltonian, which can move back the spins to be both up in the $z$ direction, thus effectively counteracting the action of the jump operators (Fig. \ref{fig:NJRJMB}).

As noted in Ref.~\cite{WisemanGambetta2008,Piilo2009}, the dynamics corresponding to this unraveling \cite{Piilo2008,Piilo2009} do not have an immediate physical representation in terms of a measurement protocol. For example, probing the bath to check if a jump occurred may destroy the information lost by the system and stored in the bath, and prevent the possibility of successively restoring such information. Nevertheless, this method provides key qualitative insights on non-Markovian dynamics, and rigorously illustrates how information back-flow from the environment to the system can be captured utilizing pure state dynamics only. Moreover we can still treat the trajectories as well defined mathematical objects, each with a quantum state that solely determines the physical properties, and a stochastic probability of realizing that trajectory. In Sec.~\ref{Sec:Diagrams}, we will finally show how, under certain conditions, the inequivalent trajectories we discuss {\it do} describe the evolution of a realistic system (albeit corresponding to a master equation that differs from the one we start from).

From these considerations, it is evident that the application of the NMQJ method as described in Ref.~\cite{Piilo2008,Piilo2009} to many-body systems is not viable. One needs a new formulation that incorporates the conceptual understanding gathered from single body problems with the non-trivial many-body dynamics, in a mathematically coherent manner: this is what we develop below.

\subsection{Many-body non-Markovian quantum jumps}\label{Sec:manybodyNMQJ}

We consider a many-body system subjected to coherent time evolution. For simplicity, we focus on a Hamiltonian dynamics, with Hamilton operator $H$: most of our reasoning also applies to stroboscopic time evolution as realized, for example, in random unitary circuits (on which we will elaborate further in the next sections). 

Let us now suppose for simplicity that only one jump operator $a$ acts on the system; the generalization to many decay channels is straightforward.

When the system does not jump, it undergoes a deterministic evolution which is governed by the Hamiltonian $H$ plus a non-Hermitian contribution arising from the back-action of the jump operator, which we can write as $H_{eff}=H-i\Delta(t)a^{\dagger}a/2$. Over a time $\delta t$ the quantum state of the system then evolves as \cite{Piilo2008,Piilo2009}
\begin{equation}\label{Eq:}
\ket{\psi(t+\delta t)}=\frac{(1-iH_{\textrm{eff}}\delta t)\ket{\psi(t)}}{||(1-iH_{\textrm{eff}}\delta t)\ket{\psi(t)}||}
\end{equation}

For book-keeping simplicity, we incorporate all these operations into an operator $\mathcal U(t,t')$ that represents the deterministic evolution between $t$ and $t'$, so that $\ket{\psi(t')}=\mathcal U(t,t')\ket{\psi(t')}$.

\begin{figure}[!t]
    \centering
    \includegraphics[width=\columnwidth]{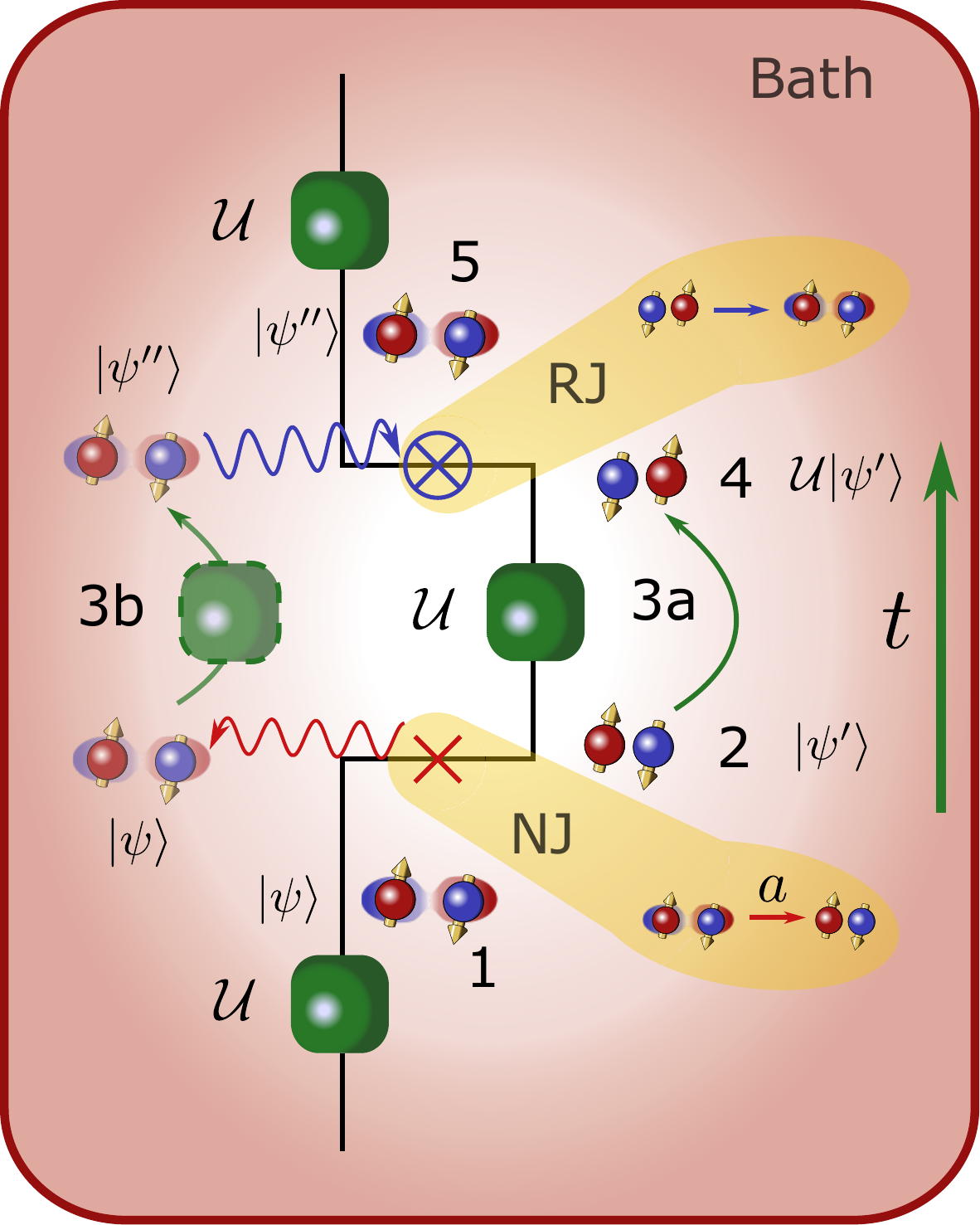}
    \caption{Example of normal jump (NJ) and reverse jump (RJ) processes in a system coupled to a non-Markovian bath. 1) A system of two spins starts in an entangled Bell pair state $(|\uparrow\downarrow\rangle+|\downarrow\uparrow\rangle)/\sqrt2$. 2) The spins undergo a NJ process (red cross) that destroys the coherence and collapses the spins onto a separable state; the information lost in the process is ``stored'' into the bath (wavy red line). 3a) The system evolves with unitary $\mathcal{U}$ (green square), for example a spin flip $\sigma_1^-\sigma_2^++\sigma_1^+\sigma_2^-$; 3b) the bath ``remembers'' the state before 2 and evolves it with $\mathcal{U}$ (transparent green square) -- in this particular case the entangled state is not changed by the spin flip. 4) The system undergoes a reverse jump (blue crossed circle): the information stored in the bath flows back (blue wavy line) into the system that 5) ends up back into the entangled state, regaining its coherence.}
    \label{fig:EvolutionNM}
\end{figure} 

As mentioned, the many-body system can jump at any time and as many times as possible, and the times at which the jump operator is applied matter since the jump operator and the unitary evolution may compete with each other.

The detailed way in which a NJ or a RJ act is shown in Fig. \ref{fig:NJRJMB} and Fig. \ref{fig:EvolutionNM}. Let us suppose that at time $t$ a jump operator $a$ is applied to the quantum state $|\psi\rangle$ of the system. The system jumps into $|\psi'\rangle=a|\psi\rangle/||a|\psi\rangle||$, but the bath retains memory of the state $|\psi\rangle$ before the jump. From $t$ to $t'$ the system then evolves with $\mathcal{U}$, $|\psi'\rangle\rightarrow\mathcal{U}|\psi'\rangle$. At time $t'$ a RJ occurs: the system does not jump back into $|\psi\rangle$, but into $|\psi''\rangle=\mathcal{U}|\psi\rangle$. In other words, the RJ brings the system back to the state it would have (unitarily) evolved into if it had never normal jumped at time $t$. The memory effect is here: the bath remembers the state of the system before the NJ and once the RJ occurs this information flows back into the system in the form of bringing it back to $|\psi''\rangle$. The operatorial definition of the RJ is $|\psi''\rangle\langle\psi'|=\mathcal{U}|\psi\rangle\langle\psi'|$; we stress that it implicitly includes the unitary evolution $\mathcal{U}$, which was absent in the single-body case.

The fact that we can reverse the last jump independently of the time it passed since its occurrence is a consequence of the infinite-time memory that we assumed for the non-Markovian bath interacting with the system. On the other end of the ``memory spectrum'', a Markovian bath has a zero time memory, so that a jump can never be erased. In the middle of the spectrum, there are baths that have a finite but non-zero time memory, so it becomes more and more unlikely to reverse a jump that occurred a long time in the past.

We note an important point: evolving the system from $t$ to $t'$ with just $\mathcal{U}$ or with the sequence $\text{RJ}\circ\mathcal{U}\circ\text{NJ}$ produces the same quantum state $|\psi''\rangle$ by definition, but along two different trajectories. These two trajectories are equivalent at time $t'$, in the sense that they correspond to the same quantum state and exhibit the same physical properties.

\begin{figure}[!t]
    \centering
    \includegraphics[width=0.9\columnwidth]{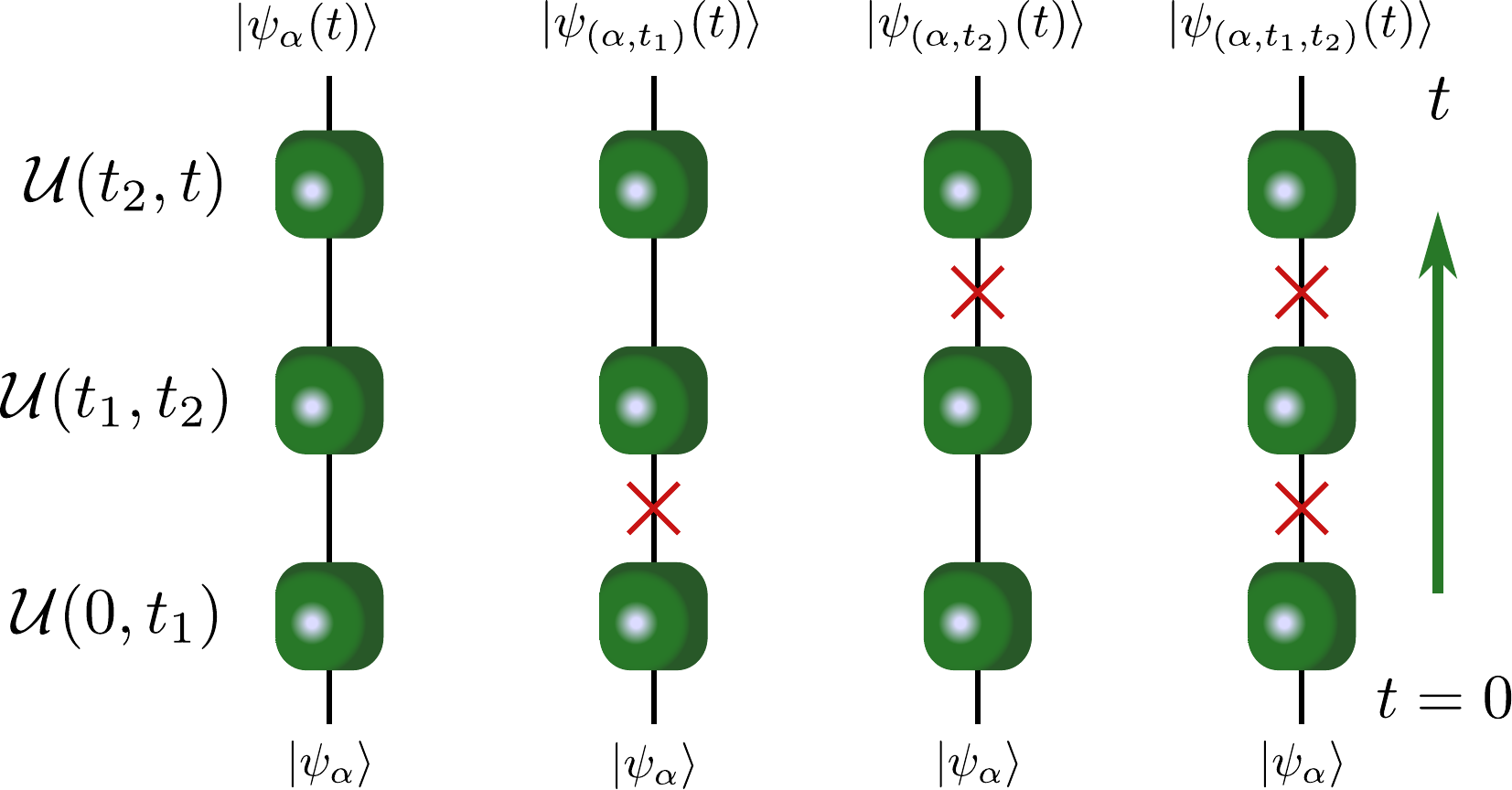}
    \caption{Different trajectory classes labeled based on the number and time of jumps. The green squares represent the periods of unitary evolution, while the red crosses represent normal quantum jumps.}
    \label{fig:alfaTraj}
\end{figure}

We can therefore group different trajectories into a \emph{class of trajectories} (labeled by the index $\alpha$) that all exhibit the same quantum state $|\psi_{\alpha}(t)\rangle$ at time $t$. We observe that $|\psi_{\alpha}(t)\rangle$ is completely specified by the initial state $|\psi\rangle$ and by the sequence of times at which normal jumps are performed without being reversed later. In other words, if we label a trajectory class with $\alpha=(t_1,t_2,...,t_n)$ (see Fig. \ref{fig:alfaTraj}), the quantum state associated to it is given by the unitary evolution, punctuated by the jump operators at the times specified by $\alpha$:
\begin{equation}\label{Eq:psialpha}
|\psi_{\alpha}(t)\rangle\equiv \frac{\mathcal U(t_n,t)a\:\mathcal U(t_{n-1},t_n)a...a\:\mathcal U(0,t_1)|\psi\rangle}{||\mathcal U a\:\mathcal U a...a\:\mathcal U |\psi\rangle||}.
\end{equation}

We remark that using this categorization into trajectory classes, the application of the NMQJ recipe is quite straightforward.

Performing a normal jump at time $t$ simply takes the state of the system from the class $\alpha=(t_1,t_2,...,t_n)$ to $\alpha'=(t_1,t_2,...,t_n,t)=(\alpha,t)$. The NJ process and its probability $p^{+}_{\alpha\rightarrow\alpha'}$ are
\begin{gather}\label{EQ:MB_MQJqs}
|\psi_{\alpha}(t)\rangle\rightarrow|\psi_{(\alpha,t)}(t)\rangle=\frac{a|\psi_{\alpha}(t)\rangle}{||a|\psi_{\alpha}(t)\rangle||};\\
\label{EQ:MB_MQJqs_prob}p^{+}_{\alpha\rightarrow\alpha'}=\Delta(t)\delta t\langle \psi_{\alpha}(t)|a^{\dagger}a|\psi_{\alpha}(t)\rangle.
\end{gather}

Conversely, performing a reverse jump from the class $\alpha=(t_1,t_2,...,t_n)=(\alpha',t_n)$ erases the last NJ performed by the system and takes it into the class $\alpha'=(t_1,t_2,...,t_{n-1})$:
\begin{gather}\label{EQ:MB_NMQJqs}
|\psi_{\alpha}(t)\rangle\rightarrow|\psi_{\alpha'}(t)\rangle;\\
\label{EQ:MB_NMQJqs_2}
\mathcal U(t_n,t)\frac{a|\psi_{\alpha'}(t_n)\rangle}{||a|\psi_{\alpha'}(t_n)\rangle||}\,\rightarrow\,\mathcal U(t_{n-1},t)|\psi_{\alpha'}(t_{n-1})\rangle.
\end{gather}
The operator describing the RJ in Eq. \eqref{EQ:MB_NMQJqs} is $|\psi_{\alpha'}(t)\rangle\langle\psi_{\alpha}(t)|$, which again includes implicitly the unitary evolution operator. The RJ in Eq. \eqref{EQ:MB_NMQJqs_2} effectively erases the jump that occurred at time $t_n$, but any trajectory that jumped at a time $t_{n-1}<t'<t$ can reverse jump from $(t_1,t_2,...,t_{n-1},t')$ back to $\alpha'$. Therefore we have to account for these possibilities in the definition of the probability to perform the reverse jump, which is given by
\begin{equation}\label{Eq:MB_NMQJprob}
p^-_{\alpha\rightarrow\alpha'}(t)=\frac{N_{\alpha'}}{\sum\limits_{t_{n-1}<t'<t}N_{(\alpha',t')}}|\Delta(t)|\delta t\langle a^{\dagger}a\rangle_{\alpha'}(t),
\end{equation}
where $\langle a^{\dagger}a\rangle_{\alpha}(t)\equiv\langle\psi_{\alpha}(t)|a^{\dagger}a|\psi_{\alpha}(t)\rangle$ and $N_{\alpha}$ is the probability for the system to be in trajectory $\alpha$. 

The RJ probability is independent of the starting state, in the sense that it is independent of the time $t'$ at which the last jump was performed: every trajectory that originates by normal jumping from the same $\alpha'$ has the same probability of performing a reverse jump back into $\alpha'$. This property may seem counter intuitive, but actually makes sense since the system does not care when the last jump occurred. For baths with a finite time memory, this is not true anymore, since the probability to reverse jump from $(\alpha',t')$ back to $\alpha'$ decreases as the time difference $t-t'$ increases. This could be quantified by introducing a memory kernel $K(t',t)$ in the fraction of Eq. \eqref{Eq:MB_NMQJprob}: $p^-_{(\alpha',t_n)\rightarrow\alpha'}(t)\sim\frac{N_{\alpha'}K(t_n,t)}{\sum\limits_{t_{n-1}<t'<t}N_{(\alpha',t')}K(t',t)}$.

We stress that the sum in the denominator is essential for the quantum jump prescription to be consistent with the master equation for the density matrix. It can be proven that averaging the dynamics described by Eqs. \eqref{Eq:MQJprob}, \eqref{EQ:MB_MQJqs}-\eqref{Eq:MB_NMQJprob} correctly recovers the master equation \eqref{Eq:NMQMasterEquation}. The calculation is tedious but straightforward if the density matrix is written as
\begin{equation}\label{Eq:rhoNonMarkov}
\rho(t)=\sum_{n=0}^{\infty}\sum_{\{\alpha=(t_1,...,t_n)\}}\frac{N_{\alpha}(t)}N|\psi_{\alpha}(t)\rangle\langle\psi_{\alpha}(t)|,
\end{equation}
where the sum over $n$ and over all the times at which the jumps can be performed exhausts all the trajectory classes generated by the evolution.

As expected, the extension of the NMQJ method to a many-body system makes the problem very hard to solve numerically. Not only the number of trajectory classes is exponential $\sim2^{N_t}$ (with $N_t$ the number of time steps in the evolution) for each decay channel, but it is not even possible to do a statistical sampling of the ensemble as in MCWF, due to the crosstalk between trajectories. Since both the individual trajectories and classes of trajectories are {\it not} independent of each other as they are in the Markovian case, all of them are needed to compute the probability of reverse jumps.

However, we observe that a class of trajectories is completely identified by $\alpha$, i.e. the times at which the \emph{normal} jumps occur. This is also true in the Markovian case, where the probability of the system ending up in the state associated to $\alpha=(t_1,t_2,...,t_n)$ can be calculated at once by multiplying the probability of jumping at times $t_1,t_2,...,t_n$ with the probability of not jumping at the other times.

In the non-Markovian case, the Markovian probability is modified -- borrowing a field theory term, we could say it gets ``dressed'' or renormalized -- by all the trajectories equivalent to $\alpha$, in which $m$ other normal jumps were performed but later cancelled out by an equal number of reverse jumps, see Fig. \ref{fig:alfaTraj} and \ref{fig:Loops}. If we can find a way to express this additional contribution we can drastically simplify the treatment of the non-Markovian dynamics. This is the topic of the next section.

\section{Diagrammatics of trajectories}\label{Sec:Diagrams}

Our goal in this section is to calculate the ``dressed'' contributions that affect the probability of realizing each state $|\psi_{\alpha}\rangle$ due to the presence of equivalent trajectories that feature a series of reverse jumps. In particular, we need to evaluate how the latter ones sum up to modify the probability of a given trajectory class. 

For the sake of concreteness, we assume that the system is evolved between $t=0$ and $t=t_f$ and consider the class $\alpha=(t_1,...,t_n)$. In the Markovian regime, there is only one trajectory contributing to this class, while in the non-Markovian regime, many trajectories contribute to the population of this class. For example any trajectory performing normal jumps at times $t_1,...,t_n$ plus any number of additional pairs of NJ plus the relative RJ are valid trajectories contributing to the population of $\alpha$. We note that the normal and reverse jumps must occur at times $t_m$ and $t_m'$ comprised between the times of two successive normal jumps in $\alpha$, i.e. such that $t_j<t_m,t_m'<t_{j+1}$, with $j=0,...,n$ and $t_0=0$ and $t_{n+1}=t_f$. These pairs of normal + reverse jumps constitute sort of ``loops" (to borrow another term from field theory) that renormalize and increase the probability of realizing the trajectory class $\alpha=(t_1,...,t_n)$, see Fig. \ref{fig:Loops}. It is the contribution of these loops that we want to calculate.

It is useful to write $\Delta(t)=\Delta_+(t)+\Delta_-(t)=\Delta_+(t)-|\Delta_-(t)|$, where $\Delta_{\pm}(t)$ is the positive/negative part of the decay rate. We also define $\Pa(t,t')$ as the conditional probability that the system is in the state labeled by $\alpha$ at time $t$ and is again found in the same state $\alpha$ at a later time $t'$. Such probability essentially corresponds to the probability that no additional normal jumps are performed between $t$ and $t'$, or that all the normal jumps performed are cancelled by an equal number of reverse jumps.

In the limit where the dynamics is Markovian, the ``bare'' probability $\Pa^{(0)}(t,t')$ is given by
\begin{equation}\label{Eq:Pazero}
\Pa^{(0)}(t,t')=\exp\left(-\int_t^{t'}d\tau\Delta_+(\tau)\langle a^{\dagger}a\rangle_{\alpha}(\tau)\right).
\end{equation}
Equation \eqref{Eq:Pazero} arises from the fact that the unitary evolution does not change the probability of the system being in a certain class, and that the product of the probabilities of performing no jumps between $t$ and $t'$ is an exponential in the continuum limit.

\subsection{No jump trajectory}

\begin{figure}[!t]
    \centering
    \includegraphics[width=0.9\columnwidth]{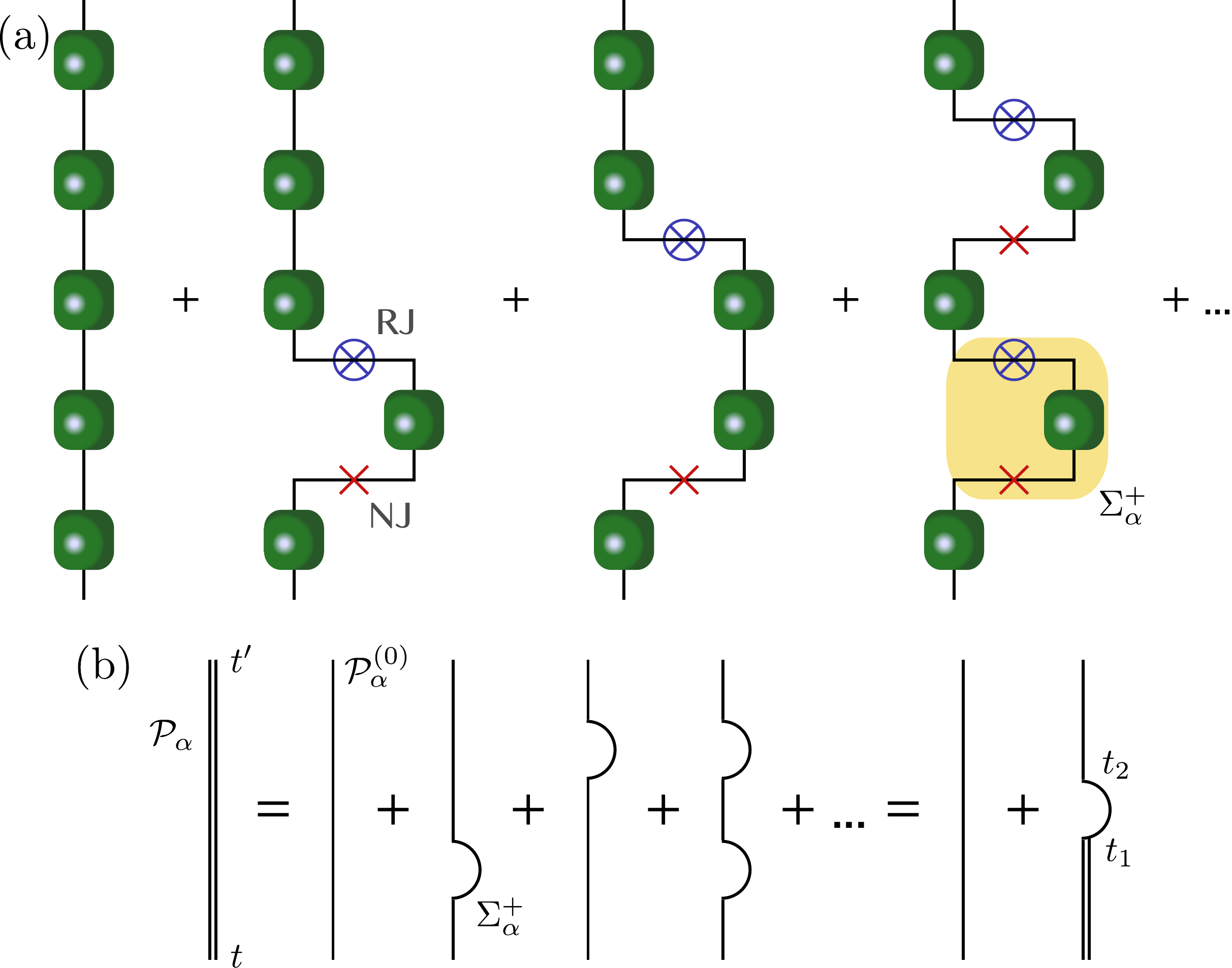}
    \caption{(a) The trajectories shown in the panel are different since they have different normal + reverse jumps sequences, but all lead to the same final quantum state. Their probabilities must then be summed. (b) Diagrammatic method to calculate the dressed propagator.}
    \label{fig:Loops}
\end{figure}

We start from the simpler case in which the class we consider is the no jump trajectory, i.e. $\alpha=\varnothing$, see Fig. \ref{fig:Loops}a. The conditional probability of staying in such trajectory is corrected (with respect to the Markovian case) only by loops of the type normal jump + reverse jump, because no reverse jump can occur first since the system has not jumped at all to begin with.

We indicate with $\Sigma^+_{\alpha}(t,t')$ the probability of performing a normal jump from the class $\alpha$ at time $t$ and then going back to $\alpha$ with a reverse jump at time $t'$. Then we can write $\mathcal{P}_{\varnothing}$ as a perturbative series in $\Sigma^{+}$:
\begin{gather}
\notag\mathcal{P}_{\varnothing}=\mathcal{P}_{\varnothing}^{(0)}+\mathcal{P}_{\varnothing}^{(0)}\circ\Sigma^+\circ\mathcal{P}_{\varnothing}^{(0)}+\\
+\mathcal{P}_{\varnothing}^{(0)}\circ\Sigma^+\circ\mathcal{P}_{\varnothing}^{(0)}\circ\Sigma^+\circ\mathcal{P}_{\varnothing}^{(0)}+...;\\
\notag\mathcal{P}_{\varnothing}=\mathcal{P}_{\varnothing}^{(0)}+(\mathcal{P}_{\varnothing}^{(0)}+\mathcal{P}_{\varnothing}^{(0)}\circ\Sigma^+\circ\mathcal{P}_{\varnothing}^{(0)}+...)\circ\Sigma^+\circ\mathcal{P}_{\varnothing}^{(0)};\\
\label{Eq:DysonNull}\mathcal{P}_{\varnothing}=\mathcal{P}_{\varnothing}^{(0)}+\mathcal{P}_{\varnothing}\circ\Sigma^+\circ\mathcal{P}_{\varnothing}^{(0)},
\end{gather}
where $\circ$ represents the convolution over all times between $t$ and $t'$, i.e. $(A\circ B)(t,t')=\int_t^{t'}(dt_1/\delta t) A(t,t_1)B(t_1,t')$ -- in the continuum limit we divide the integration over steps of length $\delta t$. The resummation formula contained in Eq. \eqref{Eq:DysonNull} is depicted graphically in Fig. \ref{fig:Loops}b.

Writing explicitly the convolutions we find
\begin{align}
\notag\mathcal{P}_{\varnothing}(t,t')&=\mathcal{P}_{\varnothing}^{(0)}(t,t')+\int_t^{t'}\frac{dt_2}{\delta t}\int_t^{t_2}\frac{dt_1}{\delta t}\mathcal{P}_{\varnothing}(t,t_1)\times\\
\label{Eq:DysonNullInt}&\times\Sigma^+_{\varnothing}(t_1,t_2)\mathcal{P}_{\varnothing}^{(0)}(t_2,t').
\end{align}
The integration limits express the causality of the jumps: the normal and reverse jumps must occur between $t$ and $t'$ at times such that $t<t_1<t_2<t'$.

It is worth noticing that, within this picture, the no-click limit takes the role of a mother trajectory: indeed, a large number of trajectories is represented by the dressed no-click case. This may suggest that the latter is particularly informative about the system dynamics, as already noted in some Markovian cases~\cite{Turkeshi2021a,turkeshi2022entanglement}.

\subsection{Generic trajectory}

Equation \eqref{Eq:DysonNullInt} can be extended to a generic conditional probability $\Pa$. In principle, there exist also reverse loops, where a reverse jump occurs first and is then followed by a normal jump. However, the action of such loops on $\Pa$ is ill-defined, in the sense that it is not an actual loop since it does not bring the system back to the same trajectory class.

To be more specific, let us assume that $\alpha=(t_1,..,t_m)$; a reverse jump at $t>t_m$ brings the system into the state labeled by $(t_1,..,t_{m-1})$ and a successive normal jump closing the reverse loop takes the system into the state labeled by $(t_1,..,t_{m-1},t')\neq\alpha$. Therefore ``reverse loops'' should not be taken into account when renormalizing $\Pa^{(0)}$ since they always bring the system into a different trajectory class \footnote{We exclude the limiting case $t'=t_n$ where the reverse and normal jumps occur at the same time.}.

It is then straightforward to generalize Eq. \eqref{Eq:DysonNullInt}
\begin{align}
\notag\Pa(t,t')=\Pa^{(0)}(t,t')+&\int\frac{dt_2}{\delta t}\frac{dt_1}{\delta t}\Pa(t,t_1)\cdot\\
\label{Eq:DysonAlphaInt}&\cdot\Sigma^+_{\alpha}(t_1,t_2)\Pa^{(0)}(t_2,t').
\end{align}

Borrowing some more terminology from field theory, we can regard the conditional probability $\Pa$ as a sort of propagator of the class $\alpha$. We observe that the ``dressed'' propagator is related to the ``bare'' propagator $\Pa^{(0)}$ by a relation very similar to the Dyson equation for the Green function of interacting systems, where the loop probability $\Sigma^+$ plays the role of the self-energy.

We write the loop probability $\Sigma^+(t_1,t_2)$ as the probability to perform a normal jump at time $t_1$, times the conditional probability to stay in the new trajectory class $(\alpha,t_1)$, times the probability to reverse jump at time $t_2$.
\begin{equation}\label{Eq:SigmaP}
\Sigma^+_{\alpha}(t_1,t_2)=p^+_{\alpha\rightarrow(\alpha,t_1)}(t_1)\mathcal{P}_{(\alpha,t_1)}(t_1,t_2)p^-_{(\alpha,t_1)\rightarrow\alpha}(t_2).
\end{equation}

We stress that in Eq. \eqref{Eq:SigmaP} the conditional probability to stay in $(\alpha,t_1)$ is ``dressed'' because we have to allow for the possibility of ``nested'' sequences of jumps, e.g. of the type NJ+NJ+RJ+RJ, in which the system jumps further away from $(\alpha,t_1)$ and then comes back to it with reverse jumps before $t_2$.

Using Eqs. \eqref{Eq:MQJprob} and \eqref{Eq:NMQJprob} we write
\begin{gather}\label{Eq:pPAlphaSigmaP}
p^+_{\alpha\rightarrow(\alpha,t_1)}(t_1)=\Delta_+(t_1)\delta t\langle a^{\dagger}a\rangle_{\alpha}(t_1);\\
\label{Eq:pMAlphaSigmaP}p^-_{(\alpha,t_1)\rightarrow\alpha}(t_2)=\frac{N_{\alpha}(t_2)|\Delta_-(t_2)|\delta t}{\int_t^{t_2}\frac{d\tau}{\delta t}N_{(\alpha,\tau)}(t_2)}\langle a^{\dagger}a\rangle_{\alpha}(t_2).
\end{gather}

In the ratio of populations of the target and sources states, we switched to the continuum limit and replaced the summation by an integration. An important point is that the integration in the denominator runs from $t$ to $t_2$. The upper limit obviously follows from causality, since we can only reverse at time $t_2$ trajectories that underwent a normal jump from $\alpha$ before $t_2$. The lower limit is a consequence of the conditional probability $\Pa(t,t')$: we condition the system to be in the state $\alpha$ at time $t$ and we have to only take into account trajectories that normal jumped from $\alpha$ after that time.

The ratio of populations is essentially a ratio of probabilities, and both numerator and denominators can factorize into the probability to be in the state $\alpha$ at time $t$ times the probability to stay in $\alpha$ (or to jump into $(\alpha,\tau)$ for the denominator):
\begin{equation}\notag
\frac{N_{\alpha}(t_2)}{\int_t^{t_2}\frac{d\tau}{\delta t}N_{(\alpha,\tau)}(t_2)}=\frac{\Pa(t,t_2)}{\int_t^{t_2}\frac{d\tau}{\delta t}\Pa(t,\tau)p^+_{\alpha\rightarrow(\alpha,\tau)}(\tau)\mathcal{P}_{(\alpha,\tau)}(\tau,t_2)}
\end{equation}

The denominator arises from the fact that the conditional probability of being in a trajectory eligible to reverse jump is the sum over all times $\tau$ between $t$ and $t_2$ of the probability $\mathcal{P}_{\alpha}(t,\tau)$ to propagate the state $\alpha$ from $t$ to $\tau$ times the probability $p^+_{\alpha\rightarrow(\alpha,\tau)}(\tau)$ of jumping at time $\tau$ times the probability $\mathcal{P}_{(\alpha,\tau)}(\tau,t_2)$ to propagate in $(\alpha,\tau)$ from $\tau$ to $t_2$.

Substituting into Eq. \eqref{Eq:pMAlphaSigmaP} and \eqref{Eq:SigmaP}, the integral in the denominator simplifies when integrating over $t_1$
\begin{gather}
\notag\int \frac{dt_1}{\delta t}\Pa(t,t_1)\Sigma^+_{\alpha}(t_1,t_2)=\int \frac{dt_1}{\delta t}\Pa(t,t_1)p^+_{\alpha\rightarrow(\alpha,t_1)}(t_1)\cdot\\
\notag\frac{\mathcal{P}_{(\alpha,t_1)}(t_1,t_2)\Pa(t,t_2)|\Delta_-(t_2)|\delta t\langle a^{\dagger}a\rangle_{\alpha}(t_2)}{\int_t^{t_2}d\tau/\delta t\Pa(t,\tau)p^+_{\alpha\rightarrow(\alpha,\tau)}(\tau)\mathcal{P}_{(\alpha,)}(\tau,t_2)}=\\
\label{Eq:PalphaSigmaP}=\Pa(t,t_2)|\Delta_-(t_2)|\delta t\langle a^{\dagger}a \rangle_{\alpha}(t_2).
\end{gather}

This result is remarkable, as after integrating over the starting time of the loop,  the specific trajectory class into which the system jumps does not matter. This is a consequence of the fact that all trajectories eligible to perform a reverse jump have the same probability to do so. Combining Eq. \eqref{Eq:pPAlphaSigmaP} and \eqref{Eq:DysonAlphaInt} we obtain
\begin{widetext}
\begin{gather}\label{Eq:Palpha}
\Pa(t,t')=\Pa^{(0)}(t,t')+\int_t^{t'}dt_2\Pa(t,t_2)|\Delta_-(t_2)|\delta t\langle\psi_{\alpha}(t_2)|a^{\dagger}a|\psi_{\alpha}(t_2)\rangle\Pa^{(0)}(t_2,t');\\
\label{Eq:PalphaExp}
\Pa(t,t')=\exp\left(-\int_t^{t'}d\tau(\Delta_+(\tau)-|\Delta_-(\tau)|)\langle\psi_{\alpha}(\tau)|a^{\dagger}a|\psi_{\alpha}(\tau)\rangle\right)=\exp\left(-\int_t^{t'}d\tau\Delta(\tau)\langle a^{\dagger}a\rangle_{\alpha}(\tau)\right).
\end{gather}
\end{widetext}

Equation \eqref{Eq:PalphaExp} is particularly telling. It implies that the regions of non-Markovianity in the decay rate renormalize the probability of staying in a certain trajectory class $\alpha$. It is also similar to the probability of staying in the excited state of a non-Markovian two level system (as obtained form solving the master equation \cite{Piilo2009}); however, it shows that this simple expression for the probability of staying in the same state is also valid for a generic many-body system, provided that the state $\ket{\psi_{\alpha}}$ associated to the label $\alpha$ changes in time according to the unitary and jump evolutions.

\subsection{Probability of a generic outcome}

We now want to calculate what is the probability of performing a certain number of normal jumps between an initial time $t=0$ and a final time $t$.

Let us start from the case of one jump, in which we go from the class $\alpha=\varnothing$ to the class $\alpha=(t_1)$ within a small time interval of width $\delta t$ and centered around time $t_1$. The probability $\mathcal{P}_{\varnothing}^{(t_1)}$ of ending up in this state is then given by:
\begin{equation*}
\mathcal{P}_{\varnothing}^{(t_1)}(0,t)=\mathcal{P}_{\varnothing}(0,t_1)\Delta_+(t_1)\delta t\langle a^{\dagger}a\rangle_{\varnothing}(t_1)\mathcal{P}_{(t_1)}(t_1,t). 
\end{equation*}

In other words the probability of the evolution realizing the outcome $(t_1)$ is given by the probability to not jump between $0$ and $t_1$, times the probability to perform a  normal jump in a $\delta t$ interval around $t_1$ times the probability to not jump between $t_1$ and $t$ and stay in the $(t_1)$ outcome.

We note that we can write $\Delta_+(t_1)$ as $\Delta(t_1)$ since normal jumps only occur in the Markovian regions of the evolution. In this sense we observe $\mathcal{P}_{\varnothing}^{(t_1)}(0,t)=(-\partial_{t_1}\mathcal{P}_{\varnothing}(t,t_1)) \delta t\mathcal{P}_{(t_1)}(t_1,t)$, or in other words the probability to jump out of the $\varnothing$ outcome at time $t_1$ is minus the time derivative of the probability to stay into that outcome.

Generalizing the above, we write the probability to jump from outcome $\alpha$ at time $t$ to outcome $(\alpha,t_1,t_2,...,t_n)$ at time $t'$ by performing $n$ jumps at times $t<t_1<t_2<...<t_n<t'$ as:
\begin{align}
\notag\mathcal{P}_{(\alpha)}^{(\alpha,t_1,...,t_n)}(t,t')&=\mathcal{P}_{(\alpha)}(t,t_1)\Delta_+(t_1)\delta t\langle a^{\dagger}a\rangle_{(\alpha)}(t_1)\times\\
\notag\times\mathcal{P}_{(\alpha,t_1)}(t_1,t_2)&\times...\times\Delta_+(t_n)\delta t\langle a^{\dagger}a\rangle_{(\alpha,t_1,...,t_{n-1})}(t_n)\times\\
\label{Eq:Palphat1tn}&\times\mathcal{P}_{(\alpha,t_1,...,t_n)}(t_n,t');\\
\notag\mathcal{P}_{(\alpha)}^{(\alpha,t_1,...,t_n)}(t,t')&=\prod_{j=0}^n\mathcal{P}_{(\alpha,t_1,...,t_j)}(t_j,t_{j+1})\times\\
\label{Eq:Palphat1tnProduct}\times\prod_{j=1}^n\Delta_+(t_j)&\delta t\langle a^{\dagger}a\rangle_{(\alpha,t_1,...,t_{j-1})}(t_j).
\end{align}
with the identifications $t_0=t$ and $t_{n+1}=t'$.

In the case of many decay channels -- each with an associated jump operator $a_s$ and decay rate $\Delta_s(t)$ -- we can write a vector of labels $\vec\alpha=(\alpha_1,\alpha_2,...,\alpha_{n_{\textrm{channels}}})$, where each $\alpha_s=(t_{s,1},t_{s,2},...t_{s,n_s})$ describes the times at which the system undergoes a jump through channel $s$. Since the channels are independent, the total propagator probability of no jump is the product of the propagator probability for each channel:
\begin{equation}\label{Eq:PalphaMultiChan}
\mathcal{P}_{\vec\alpha}(t,t')=\exp\left(-\int_t^{t'}d\tau\sum_s\Delta_s(\tau)\langle a_s^{\dagger}a_s\rangle_{\vec\alpha}(\tau)\right).
\end{equation}

A similar generalization of Eq. \eqref{Eq:Palphat1tnProduct} can be written down.

\subsection{Advantages and limitations of the diagrammatic renormalization method}

In this section we have shown that it is possible to obtain an analytic expression for the probability of a non-Markovian system realizing a certain sequence $\alpha$ of normal quantum jumps and ending up in the corresponding state $\ket{\psi_{\alpha}}$. This is a remarkable result, as it generalizes known results for the dynamics of Markovian systems to non-Markovian many-body systems.

However, there are some limitations to the applicability of this formula. One limit is that the results we presented are technically exact in the limit in which the system is able to jump an infinite number of times. In fact, in writing the expression for the $\Sigma^+$ loops and their corrections, we assumed that the state in which the system jumps is again eligible to jump itself, which is not the case if the system is only able to jump a finite number of times.

The comparison with the extreme example, in which the system may only jump once, shows that our equations correctly predict the probability to perform zero jumps, see Eq. (B5) in Ref. \cite{Piilo2009}, but differ from the probability of performing one jump, see Eq. (B6) in \cite{Piilo2009}. However, this is not a fatal issue, as our analytic results are more and more a good approximation as the maximum number of jumps increases, and are essentially indistinguishable from the exact results when considering large enough systems and long enough time evolutions.

Another practical issue is that applying equations \eqref{Eq:Palphat1tn}-\eqref{Eq:PalphaMultiChan} to real system still generally requires the knowledge of the quantum state of the system $\ket{\psi_{\alpha}(t)}$, which implies solving the dynamics of a many-body system, which is exponentially complex in the system size. Note that in a usual non-Markovian setting, the simulation of all possible trajectories is required, meaning the complexity is still exponential in the system size \emph{and} in the evolution time. However, there are some special cases in which the physics of a system can be studied without needing to know the quantum state of the system at all times; one of them is the case of the mapping of random unitary circuits into a statistical model \cite{Jian2020:RUC}, which we analyze in detail in the next section.

\section{Non-Markovian measurement induced transition}\label{Sec:nMEntanglement}

In this section we apply the results obtained in Section \ref{Sec:Diagrams} to investigate the dynamics of the entanglement and the transition induced by measurements in non-Markovian systems. The entanglement transition has been studied in many different systems, including random Haar \cite{Li18,Li2019, Nahum2017,Nahum18,Noh20,Napp20} and Clifford circuits \cite{Li18,Li2019,Zhou20,Lunt21,Weinstein2022:MIPT,Kelly2022:MIPT}, free fermions \cite{Cao19,Alberton21,Coppola2022,Gal22,Ladewig22}, Ising chains \cite{Li2019,Lang20,Turkeshi2021a,Piccitto22,turkeshi2022entanglement}, stabilizer circuits \cite{Gullans2020,Ippoliti21,Lavasani21,Sang21,Sierant22,Sharma2022,Klocke22}, etc.

We specialize to the case of random unitary (Haar) circuits for a number of reasons. They have been extensively studied in the literature, so there is an abundance of study cases to use for comparison; moreover, random circuits can be mapped to a classical Potts model on which either analytical or Monte Carlo calculations can be performed. And most importantly, the measurement protocol usually implemented on such circuits is such that the exponent in Eq. \eqref{Eq:PalphaMultiChan} simplifies and does not contain the quantum state of the system, greatly simplifying further analytical calculations.

\subsection{Random Unitary Circuits}

\begin{figure}[!t]
    \centering
    \includegraphics[width=1.02\columnwidth]{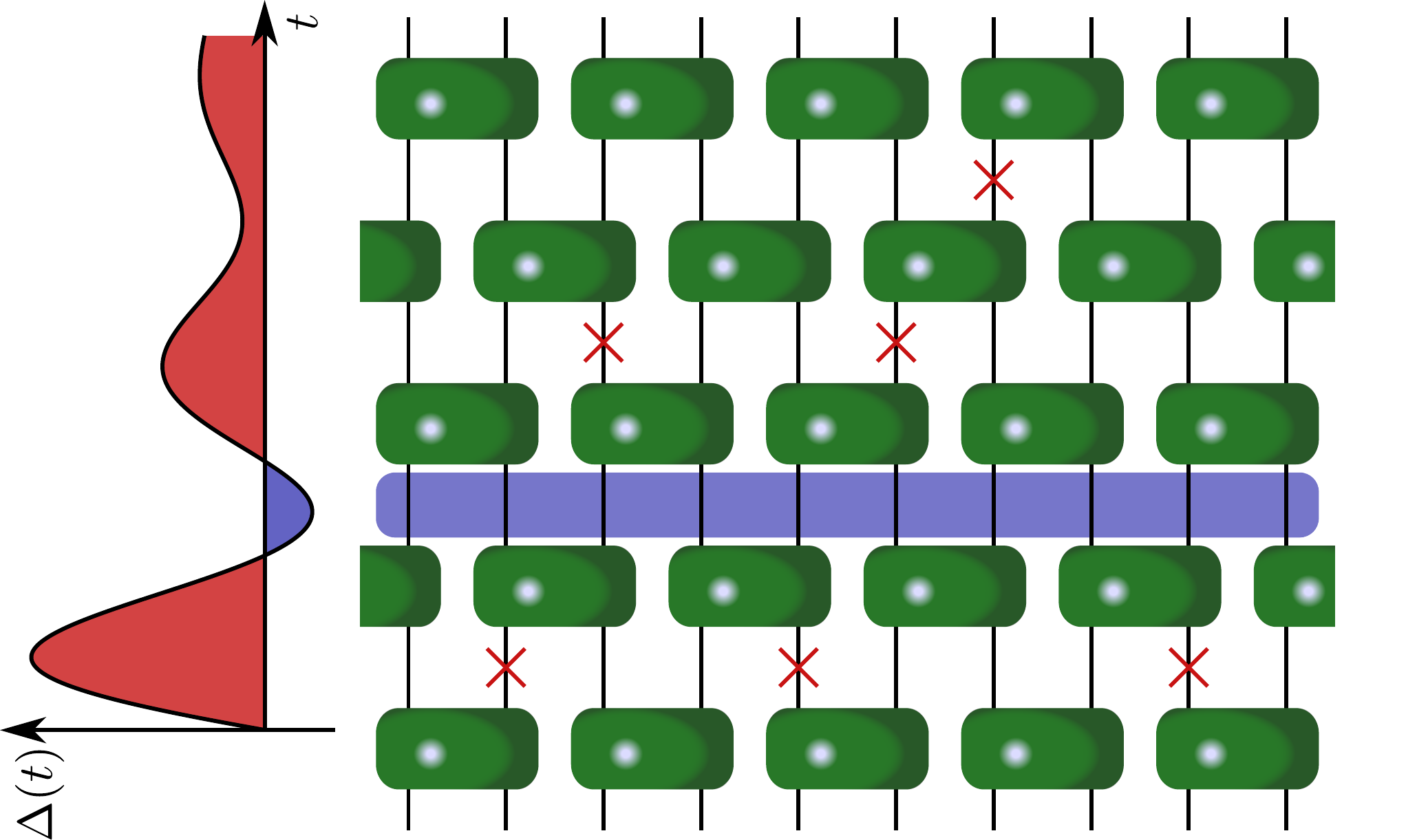}
    \caption{Diagram of a non-Markovian random unitary circuit. Layers of two-qudits unitary gates (green rectangles) alternate with layers of local random measurements (red crosses). Whenever the decay rate becomes negative (blue shaded region), no normal jump measurements are allowed; this corresponds to a ``frozen'' layer where the Potts spins behave ferromagnetically.}
    \label{fig:NMcircuit}
\end{figure}

We consider a random unitary circuit similar to the model studied in Ref. \cite{Jian2020:RUC}. The system is composed of $L$ q-dits, i.e. spins with a $d$-dimensional Hilbert space. Every time step the q-dits evolve according to random unitary gates coupling the odd or even pairs alternatively, and then undergo random local measurements, see Fig. \ref{fig:NMcircuit}. The unitary evolution does not affect the probability of being in a certain sequence $\alpha$ of quantum jumps in any way other than changing the state of the system.

Similarly to Eq. \eqref{Eq:psialpha}, we describe the state of the system at time $t_i$ by a sequence of random unitaries $\mathcal{U}$ and local normal quantum jumps $a$ applied to the initial state:
\begin{gather}\label{psiC}
\ket{\psi(t)}=\frac{\mathcal{C}(t)\ket{\psi}}{||\mathcal{C}(t)\ket{\psi}||};\\
\label{CircuitC}\mathcal{C}(t)=\mathcal U(t_n,t)a\:\mathcal U(t_{n-1},t_n)a...a\:\mathcal U(0,t_1)
\end{gather}
where $\mathcal{C}(t)$ is called circuit operator, $\alpha=(t_1,...,t_n)$ and with the obvious generalization to multiple channels of decay.

The probability ${P}_{\mathcal{C}}$ of realizing a particular $\mathcal{C}$ depends on the probability $P_\mathcal{C}^{\mathcal{U}}$ associated to the random unitaries and the probability $P_\mathcal{C}^{\mathcal{M}}$ of performing the sequence of normal jumps specified by $\mathcal{C}$. Note that $P_\mathcal{C}^{\mathcal{U}}$ and $P_\mathcal{C}^{\mathcal{M}}$ are independent, so we may only focus on the probability associated to the quantum jumps, which is essentially a discretized version of Eq. \eqref{Eq:Palphat1tn}.

We now specify the protocol for the measurement: we choose Kraus operators that have equal weight and that constitute a resolution of the identity. For example, for each site we may have $d$ quantum channels, each corresponding to a projector on every state of the local Hilbert space $a_s=\ket{s}\bra{s}$ (for $s=1,...,d$); alternatively, we may have a continuous set of jump operators obtained by transforming with random unitaries the projector on one of the states $a_s=\ket{s}\bra{s}$. We only require that each jump operator in this set has an equal weight, i.e. $\Delta_s=\Delta$. This is a crucial assumption, since it simplifies the sum over the decay channels in Eq. \eqref{Eq:PalphaMultiChan}:
\begin{gather}\label{Eq:SumChannels}
\sum_s\Delta_s(\tau)\langle\psi(\tau)|s\rangle\langle s|s\rangle\langle s|\psi(\tau)\rangle=\\
\notag=\Delta(\tau)\langle \psi(\tau)|\sum_s|s\rangle\langle s|\psi(\tau)\rangle=\Delta(\tau)
\end{gather}
since $\sum_s|s\rangle\langle s|=\mathbb{1}$. With this simplification, the probability of no jump becomes independent of the quantum state of the system, and it is possible to calculate it without having to study the many-body dynamics of the system.

\subsection{Probability of a circuit realization}

We discretize the evolution: for any measurement time $t_i$ we define $p_i=\Delta(t_i)\delta t$. During the Markovian regions $\Delta(t_i)>0$ this is a real probability of performing a jump. During the non-Markovian regions $p_i$ is negative and is not a physical probability, but still makes sense with the interpretation that when $p_i<0$ there is no normal jump and there is an increase of the weight associated to the no jump trajectory.

Indeed, the probability of performing no jumps on a certain site from time $t_i$ to time $t_{i'}$ is obtained by discretizing the propagator probability \eqref{Eq:PalphaMultiChan}:
\begin{equation}\label{Eq:PaDiscrete}
\Pa(t_i,t_{i'})=\exp\left(-\int_{t_i}^{t_{i'}}d\tau\Delta(\tau)\right)\rightarrow\prod_{j=i}^{i'}(1-p_j).
\end{equation}
i.e. the probability to not perform any normal jump is given by the probability to not undergo jumps at any of the intermediate times. Regions of non-Markovianity increase this probability, which is intuitively and formally correct, since non-Markovianity makes information flow back into the no jump outcome.

We now consider the probability of performing a normal jump and split it into two parts: one associated to the probability of performing a jump and one associated to the weight of the trajectory where the jump is $a_s=\ket{s}\bra{s}$:
\begin{align}
\notag\Delta_+(t_i)\delta t&\langle a_s^{\dagger}a_s\rangle(t_i)=p_i||a_s\ket{\psi(t_i)}||^2=\\
\label{Eq:PjumpDiscr}&=p_i\frac{||a_s\mathcal{C}(t_i^-)\ket{\psi}||^2}{||\mathcal{C}(t_i^-)\ket{\psi}||^2}=
p_i\frac{||\mathcal{C}(t_i^+)\ket{\psi}||^2}{||\mathcal{C}(t_i^-)\ket{\psi}||^2},
\end{align}
where $\mathcal{C}(t_i^\pm)$ is the circuit operator immediately after/before the normal jump. We have used that $\ket{\psi(t_i)}=\mathcal{C}(t_i^-)\ket{\psi}/||\mathcal{C}(t_i^-)\ket{\psi}||$ and $a_s\mathcal{C}(t_i^-)=\mathcal{C}(t_{i}^+)$. We have split the probability associated to the decay rate, i.e. $p_i=\Delta(t_i)\delta t$, from the probabilities associated to the weight of the trajectory, i.e $\mathcal{C}(t)\ket{\psi}$. We are now able to write the discretized form of Eq. \eqref{Eq:Palphat1tnProduct}.

We assume that the system evolves from time $t_0=0$ to time $t_m=t$, and that the circuit operator $\mathcal{C}$ describes $n$ normal jumps at times $t_{i_1},...,t_{i_n}$, no jumps at times $t_j\neq t_{i_1},...,t_{i_n}$ and a certain realization of random unitaries in between described by the probability $P^\mathcal{U}_{\mathcal{C}}$. We can then write the probability of realizing $\mathcal{C}$ associated to the quantum jumps as
\begin{equation}\label{PCnotSimplified}
P_{\mathcal{C}}=P^\mathcal{U}_{\mathcal{C}}\prod_{i\neq i_1,...,i_n}(1-p_i)\prod_{a=1}^np_{i_a}\frac{||\mathcal{C}(t_{i_a}^+)\ket{\psi}||^2}{||\mathcal{C}(t_{i_a}^-)\ket{\psi}||^2}.
\end{equation}

The circuit operators between two successive jump times only differ by a sequence of unitary operators: $\mathcal{C}(t_{i_{a+1}}^-)=\left(\prod_{j=i_a+1}^{i_{a+1}}\mathcal{U}_j\right)\mathcal{C}(t_{i_{a}}^+)$. Since the unitaries do not change the norm of the state we have $||\mathcal{C}(t_{i_{a+1}}^-)\ket{\psi}||=||\mathcal{C}(t_{i_a}^+)\ket{\psi}||$. Therefore, the product of the ratio of the norms simplifies
\begin{equation*}
\prod_{a=1}^n\frac{||\mathcal{C}(t_{i_a}^+)\ket{\psi}||}{||\mathcal{C}(t_{i_a}^-)\ket{\psi}||}=\prod_{a=1}^n\frac{||\mathcal{C}(t_{i_{a+1}}^-)\ket{\psi}||}{||\mathcal{C}(t_{i_a}^-)\ket{\psi}||}=\frac{||\mathcal{C}(t_{i_{n+1}}^-)\ket{\psi}||^2}{||\mathcal{C}(t_{i_1}^-)\ket{\psi}||^2},
\end{equation*}
which reduces to $||\mathcal{C}(t)\ket{\psi}||^2$ because $||\mathcal{C}(t_{i_{n+1}}^-)\ket{\psi}||^2=||\mathcal{C}(t)\ket{\psi}||^2$ and $||\mathcal{C}(t_{i_1}^-)\ket{\psi}||^2=1$.

Therefore we write
\begin{gather}\label{Eq:PC}
P_{\mathcal{C}}=||\mathcal{C}\ket{\psi}||^2P^\mathcal{U}_{\mathcal{C}}\prod_{a=1}^np_{i_a}\prod_{i\neq i_1,...,i_n}(1-p_i);\\
\notag P_{\mathcal{C}}=||\mathcal{C}\ket{\psi}||^2P^\mathcal{U}_{\mathcal{C}}P^\mathcal{M}_{\mathcal{C}};\\
\label{Eq:PC_Multi}
P^{\mathcal{M}}_{\mathcal{C}}=\prod_{l=1}^L\left(\prod_{a_l=1}^{n_l}p^l_{i_{a_l}}\prod_{i\neq i_{1},...,i_{n_l}}(1-p^l_{i})\right).
\end{gather}

Equation \eqref{Eq:PC_Multi} is the generalization to the multiple sites case, with $p^l_i$ the probability for a quantum jump to occur at site $l$ at time $t_i$.

The total probability is $P_{\mathcal{C}}=||\mathcal{C}\ket{\psi}||^2P^\mathcal{U}_{\mathcal{C}}\mathcal{P}_{\mathcal{C}}^{\mathcal{M}}$. The first factor is the norm of the state after applying the circuit operator, and accounts for the probability of the state to be eligible to perform a jump. The second factor is the probability of a specific realization of random unitaries.

The third factor in the product is associated to the weight for the random measurements. This crucial factorization allows us to separate the contributions that depend on the quantum state (and that thus require exponentially complex numerical calculations) from the contributions that depend on the decay rates of the quantum channels. In other words, the average over the random measurements factorizes -- as in the Markovian case -- as the product of the averages over measurements for each time of the evolution and for each site.

We reiterate that one important difference is that for the non-Markovian regions the probability to perform a jump is zero (since no normal jumps can be performed). This is a consequence of the fact that the state of the system is not affected by reverse jumps, in the sense that the final quantum state is only determined by the sequence of normal jumps; the system only cares about reverse jumps to the extent that they renormalize the probability of the system being in a certain quantum state. Indeed, a second difference of the non-Markovian regions is that the probability to not perform any jump is larger than one -- meaning a renormalization of the no jump weight. While the meaning of this probability being greater than one is apparently not very physical, this recipe is formally correct and can be employed to map the system to a classical Potts model amenable to Monte Carlo simulations.

\subsection{Mapping to a Potts model}

We use the formal mapping machinery of Ref. \cite{Jian2020:RUC}. The $n$-th Renyi entanglement entropy of a partition $A$ of the system is expressed in terms of the free energy $F$ of a replicated system where $Q$ replicas live on each site:
\begin{equation}\label{Eq:SnA}
S_n^A=\frac{n}{n-1}\lim_{Q\rightarrow1}\frac{F_A-F_0}{Q-1}
\end{equation}
where the free energy is calculated averaging over $P_{\mathcal{C}}^{\mathcal{U}}P_{\mathcal{C}}^{\mathcal{M}}$;
\begin{equation}\label{Eq:FreeEnergy}
F=-\ln\mathcal{Z}=-\ln\sum_{\mathcal{C}}P_{\mathcal{C}}^{\mathcal{U}}P_{\mathcal{C}}^{\mathcal{M}}.
\end{equation}
$F_A$ is calculated for boundary conditions (in the physical and replica space) dictated by the partition $A$ and the order $n$ of the Renyi entropy, while $F_0$ corresponds to a replica system with no partition of the system. Without going too much into the details of the mapping (which are discussed extensively in the literature \cite{Zhou2019:nahumRUC,Jian2020:RUC}), the 1+1 quantum model is mapped onto a 2+0 dimensional classical model, where each site is associated to a permutation of the replicas. Thus the classical model is essentially a $Q!$-states Potts model, where neighboring Potts spins are coupled via the unitary gates or via the measurements.

We can split the sum over unitaries and over measurements in Eq. \eqref{Eq:FreeEnergy}. The sum over the unitaries immediately factorizes into the sum over unitaries for each site \cite{Zhou2019:nahumRUC,Jian2020:RUC}, yielding terms proportional to the Weingarten functions. The sum over the random measurements also factorizes as (Eq. (13) of \cite{Jian2020:RUC})
\begin{gather}\label{Eq:AverageM}
\sum_{\mathcal{C}}P_{\mathcal{C}}^{\mathcal{M}}=\prod_{\langle j,l\rangle}\sum_{g_j,g_l\in S_Q}W_p(g^{-1}_j(t_i)g_l(t_{i+1})).
\end{gather}
where $S_Q$ is the set of permutations of $Q$ elements. The weight $W_p$ is the average over the possible outcomes of a random jump occurring on site $j$ (associated to a Potts spin $g_j$) at time $t_i$ and coupling to the next neighbor site $l$ at time $t_{i+1}$ (with associated Potts spin $g_l$). 

The expression of $W_p$ depends on the local Hilbert space dimension $d$, on the probability of jumping $p_i$ and on whether the set of normal jump operators is a discrete -- i.e. $\mathcal{M}_p=\{\mathbb{1},a_1,...,a_d\}$ with $a_s=|s\rangle\langle s|$ and weights $\{1-p,p,...,p\}$ -- or a continuous set of randomly generated projectors $\mathcal{M}_{p} =\{\mathbb{I}\} 	\cup \{\sqrt{d} a_{U} | U \in U(d) \}$, with $a_{U} = U^{\dag}a_1U$ and $U$ a random unitary matrix. For computational convenience we focus on the second option and find
\begin{gather}\label{Eq:ExpectationContMar}
W_p(g)=(1-p_i)d^{|g|}+p_id^Q\qquad p_i\geq0;\\
\label{Eq:ExpectationContNonMar}W_p(g)=(1-p_i)d^{|g|}\qquad p_i<0
\end{gather}
where $|g|\leq Q$ is the number of cycles in the permutation $g$.

The average over unitaries and measurements can be written in terms of the product over triangular plaquettes of the integrated weight $J_p(g_i,g_j;g_k)$.
\begin{gather}\label{Eq:ZA} \sum_{\mathcal{C}}P_{\mathcal{C}}^{\mathcal{M}}P_{\mathcal{C}}^{\mathcal{U}} = \sum_{\{ g_i \in S_Q\} } \prod_{\langle ijk \rangle \in \triangledown}
    J_p(g_i,g_j;g_k);\\
\notag J_p(g_i,g_j;g_k) = \sum_{g_l \in S_Q} W_{p}(g^{-1}_ig_l) W_{p}(g^{-1}_jg_l)\text{Wg}_{d^2}(g^{-1}_lg_k),
\end{gather}
where $\text{Wg}_{d^2}$ is the Weingarten function that expresses the weight associated to the random unitary evolution. In other terms, we integrate out the contribution of the unitary evolution in order to obtain a reduced average over the random measurements only.

We remark that the factorizations Eqs. \eqref{Eq:AverageM} and \eqref{Eq:ZA} only work if we want to calculate the average of operators local in time and space; however this is the case for the entanglement entropy.
\begin{figure}[!t]
    \centering
    \includegraphics[width=\columnwidth]{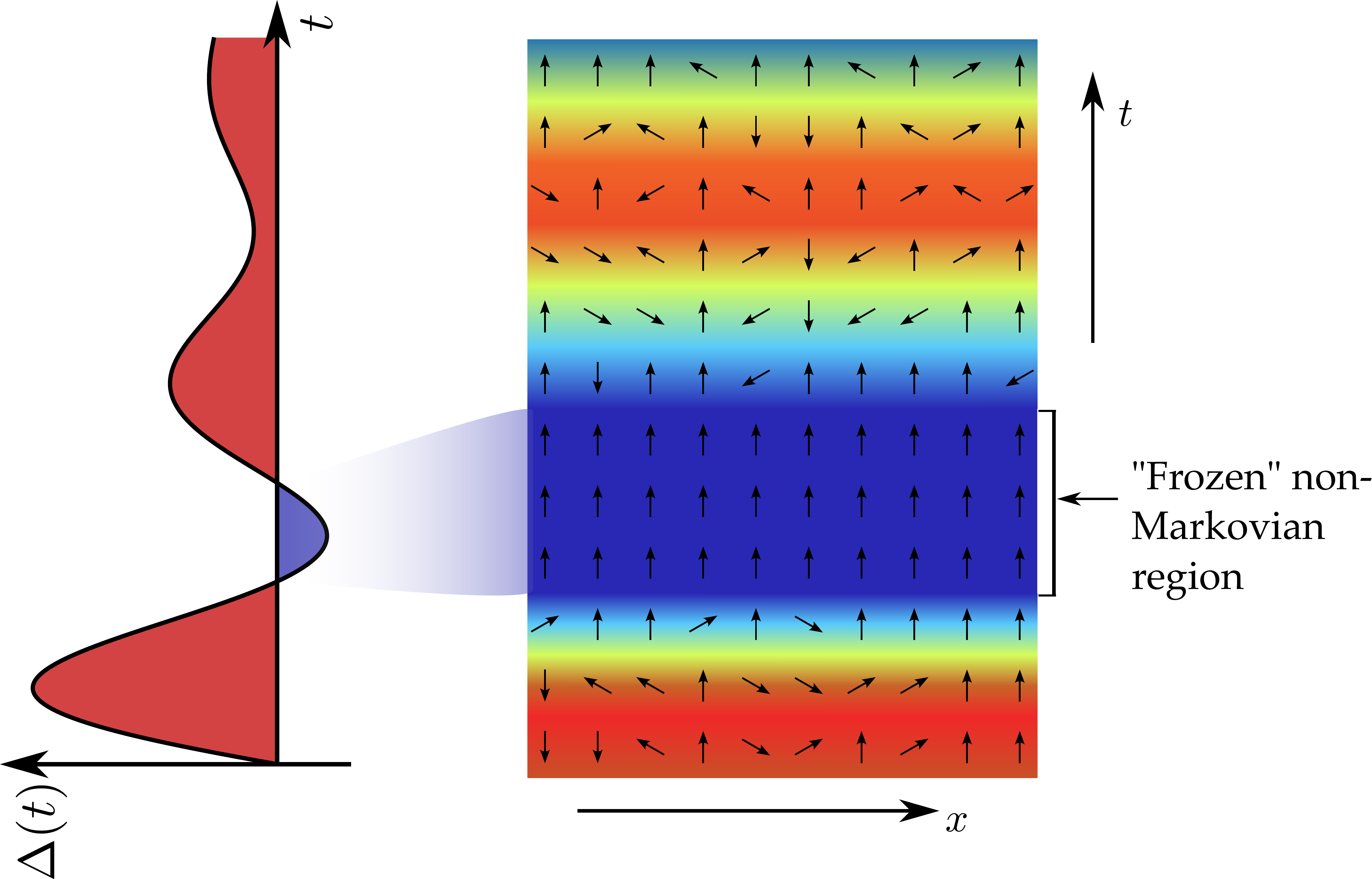}
    \caption{Sketch of the time-dependent rate $\Delta(t)$ with the non-Markovian region shaded in blue. This region corresponds to a ferromagnetic region in the Potts model, (shaded in blue on the right). The Markovian regions (shaded in red in the left plot) allow for paramagnetic regions in the Potts model, with neighboring spins not aligned.}
    \label{fig:PottsScheme}
\end{figure}

In the large $d$ limit we have \cite{Jian2020:RUC}
\begin{gather}\label{Eq:JpEik}
    J_p(g_i,g_j;g_k) = e^{-E_i(g^{-1}_ig_k)}e^{-E_j(g^{-1}_jg_k)};\\
 \label{Eq:Eik}   E_i(g)=-\ln\left((1-p_i)\left(\delta_{g}+\frac{\delta'_{g}}d\right) +\theta_{p_i}p_i\right),
\end{gather}
where $\theta_{p_i}$ is equal to $1$ for Markovian regions $p_i>0$ and equal to $0$ for non-Markovian regions $p_i<0$, and $\delta_g$ ($\delta'_g$) is one if $g$ is the identity (a transposition) and zero otherwise.

Equation \eqref{Eq:Eik} is the basis for our subsequent analysis. Given any decay rate $\Delta(t)$ we can compute the inhomogeneous couplings between different sites on the Potts model. This allows us to perform numerical Monte Carlo simulations as well as do a qualitative analysis of the effect of non-Markovianity on the entanglement transition.

In particular we know from the Markovian case that low $p$ are associated to a ferromagnetic configurations of the Potts spins and to a volume law scaling of the entanglement, i.e. a linear dependence of $F_A-F_0$ with the size $l_A$ of $A$. In fact, from Eq. \eqref{Eq:Eik} we observe that if spins are aligned (i.e. $g_i^{-1}g_k$ is the identity) the energy $E_i$ vanishes while it is approximately $E_i=-\ln p$ when they are different; thus, at low $p$, spins tend to align while at larger $p$ paramagnetic configurations with the spins aligned in random directions are possible.

In the non-Markovian regions, the energy is $-\ln(1-p_i)<0$ for aligned spins and infinite otherwise (technically the energy is finite due to $\mathcal{O}(1/d^a)$ corrections, but still  very large); therefore regions of non-Markovianity are essentially strips of ``frozen'' spins all aligned to each other (see Fig. \ref{fig:PottsScheme}), which means that they favor a volume law entanglement. This is equivalent to saying that memory effects do in fact slow down the effect of noise, and strengthen the role of coherent dynamics.

\subsection{Monte Carlo simulations}

In this section we show the results of Monte Carlo simulations.

In Eq. \eqref{Eq:SnA} the number of replicas can be expressed as $Q=nm+1$, with $m\rightarrow0$ an integer; we notice that in practice we cannot actually use the limit $m=0$ because otherwise the numerics would be trivial. Similarly, we cannot use $n=1$ because we would not be calculating an entanglement entropy (or in other words the boundary conditions would be trivial). Therefore the lowest number of replicas we can consider is $Q=3$ ($n=2$ and $m=1$), corresponding to a Potts model with six states.
\begin{figure*}[!t]
    \centering
    \includegraphics[width=\textwidth]{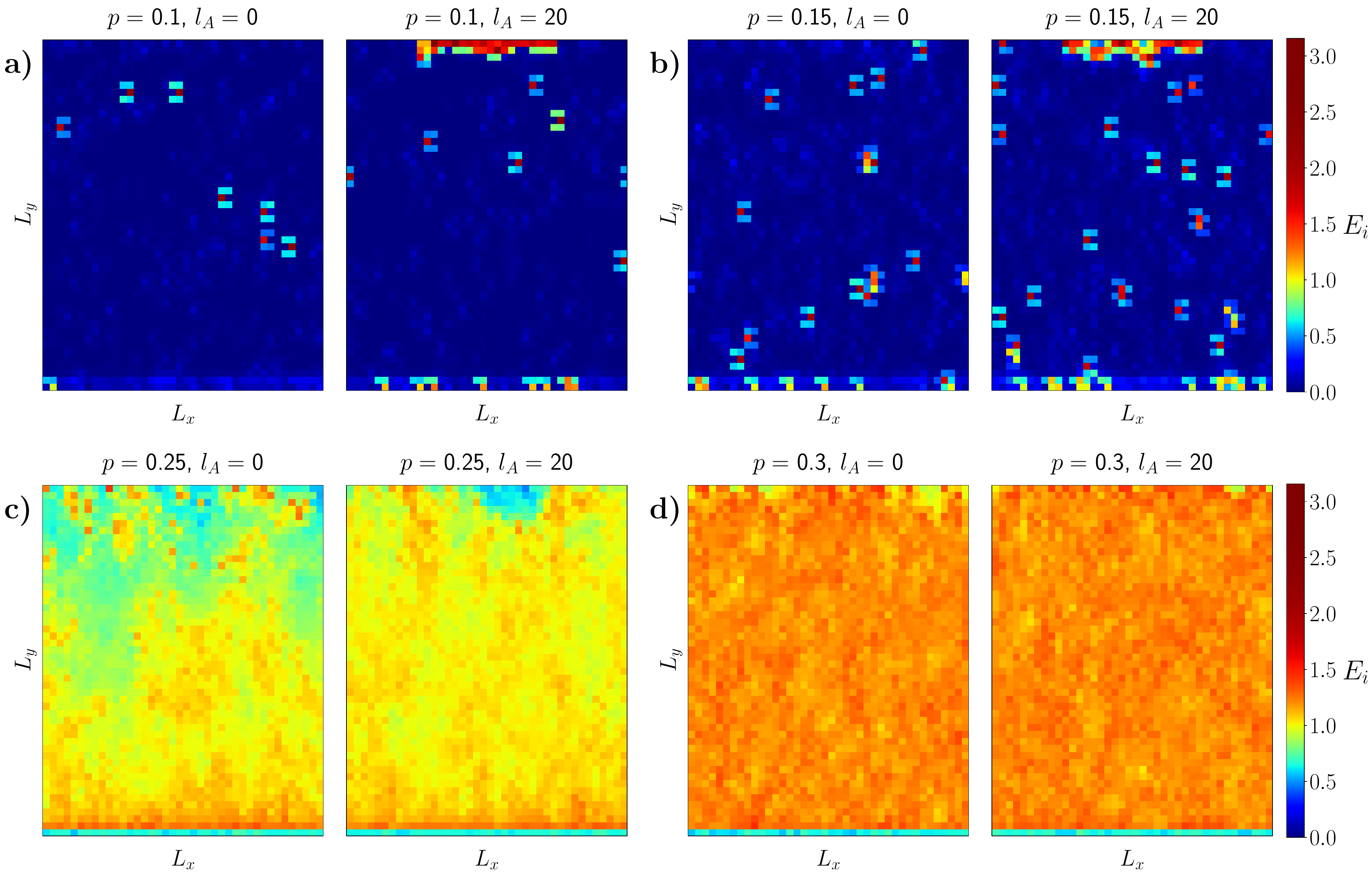}
    \caption{Colormap of the energy $E_i$ of each spin, calculated including the next neighbors contribution and the boundary contribution for the spins at the top of the chain. The size of the Monte Carlo system is $L_x=40$ and $L_y=50$. Each panel shows $l_A=0$ or $l_A=20$ for (a) $p=0.1$, (b) $p=0.15$, (c) $p=0.25$, (d) $p=0.3$. Lower energy corresponds to aligned spins, i.e. ferromagnetic regions, while larger energy corresponds to paramagnetic spins. The labels $L_y$ in the vertical direction correspond to the time direction in the physical system. a-b) The system is completely ferromagnetic and the energy cost of having a boundary is clearly visible at the top of the right plot, but small paramagnetic droplets are forming. c) The system is switching to a paramagnetic phase and the higher energy cost of the boundary is barely visible. d) The system is entirely paramagnetic and the energy cost of the boundary vanishes. Note that the bottom boundary has lower energy because it has less next neighbors than the spins in the bulk.}
    \label{fig:CMapHom}
\end{figure*}

We perform Monte Carlo simulations on a lattice of size $L_x=40$ and $L_y=50$ sites (corresponding to a time evolution of $L_y$ time periods), with periodic boundary conditions in the $x$ direction, and boundary conditions at the top in the vertical direction dictated by the value of the partition size $l_A$.

In this model the identity permutation is $(0)(1)(2)$, while the transposition dictated by the boundary conditions is $(01)(2)$ because ($n=2$ and $m=1$). We can naturally map the permutations onto spin states: $\{(0)(1)(2),(0)(12),(01)(2),(021),(012),(02)(1)\}\rightarrow s=\{0,1,2,3,4,5\}$. Therefore the boundary conditions of a partition of size $l_A$ are given by $l_A$ sites occupied by the spin $s=2$ and $L_x-l_A$ sites with the spin $s=0$. We choose the partition to be centered in the middle of the boundary.
\begin{figure}[!t]
    \centering
    \includegraphics[width=0.96\columnwidth]{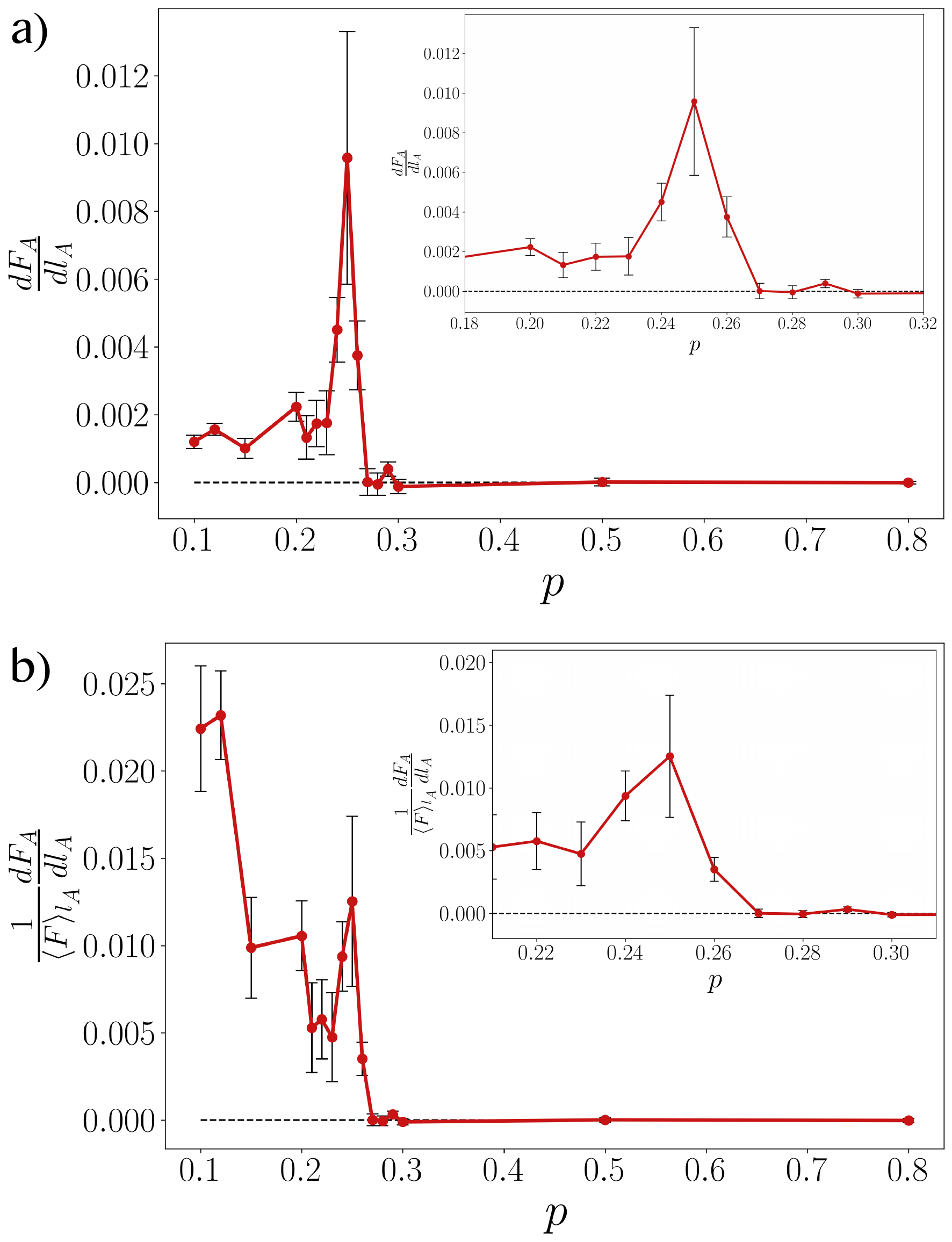}
    \caption{Behavior of the slope of the normalized energy $F_A/\langle F_A\rangle_{l_A}$ with respect to $l_A$ as function of $p$. The inset shows the slope of $F_A$ as function of $p$. The Monte Carlo calculations were performed for $L_x=40$, $L_y=50$ and $l_A=0,2,4,...,20$.}
    \label{fig:FitHom}
\end{figure}

We employ the Wolff cluster algorithm \cite{Wolff1989,Kent2018:Wolff} for the update of the lattice configuration. The probability of adding a site to the cluster built by the Wolff algorithm is the usual one, and based on the interaction energy with the neighbors. Whenever a site on the top vertical boundary is added to the cluster, we add its interaction energy with the fixed boundary to the boundary energy $E_b$. When the cluster is built, we update it with probability $\text{min}(1,e^{-E_b})$, in order to take into account the fact that configurations that have a high interaction energy with the cluster are less probable.

We first thermalize the lattice by updating it with $N_{\textrm{therm}}=25000$ Wolff steps. To avoid autocorrelations, we then sample the configuration of the lattice every $N_{\rm{sample}}=50$ steps and calculate the observables of interest.

\paragraph{Markovian Monte Carlo} -- We start by considering the Markovian case. For different values of the probability $p$ we consider different sizes of the boundary ranging from $l_A=0$ to $l_A=L_x/2$, and for each calculate the free energy $F(l_A)$. We may also consider the local energy of each lattice sites due to the interaction with its next neighbors (and with the boundary). Since aligned spins have zero interaction energy while spins oriented in different directions contribute an energy $E_i\sim-\ln p$, we are immediately able to identify the ferromagnetic and the paramagnetic regions by plotting a color map of the local energy.

We notice that at $p=p_c\approx0.25$, the phase of the system changes from ferromagnetic to paramagnetic, as indicated by the increase in energy over the entire lattice, see Fig. \ref{fig:CMapHom}. Simultaneously, the energy cost of having boundary conditions with $l_A\neq0$ is large -- Fig. \ref{fig:CMapHom}a-b -- at low $p$, decreases significantly for $p\rightarrow p_c$ (Fig. \ref{fig:CMapHom}c) and becomes negligible above the critical probability, as in a paramagnetic phase the boundary can be accommodated with very little increase in energy, see Fig. \ref{fig:CMapHom}d.

The transition is also observed by performing a linear fit of the total energy $F_A=F(p,l_A)$ as function of $l_A$ and plotting the behavior of the slope $dF_A/dl_A$ as function of $p$. When $dF_A/dl_A\neq0$ the energy of the Potts model, and thus the entanglement entropy of the circuit, scales with the size of the partition subsystem, i.e. it obeys a volume law; when $dF_A/dl_A=0$, the circuit entanglement is in an area law.

We observe a sharp transition of the slope from non zero values for $p<p_c$ to very small values for $p>p_c$, see Fig. \ref{fig:FitHom}a. This is also the case if we perform the linear fit analysis on the energy normalized to its average value over $l_A$ at fixed $p$ -- i.e. $F_A/\langle F_A\rangle_{l_A}$ -- see Fig. \ref{fig:FitHom}b. This procedure may be necessary to avoid large fluctuations, since at high $p$ the total energy becomes large and analyzing the normalized energy may be more sensible. From both fitting methods we find $p_c\approx0.25$.

We also observe a local peak of $dF_A/dl_A$ and $\frac1{\langle F_A\rangle_{l_A}}\frac{dF_A}{dl_A}$ at $p=p_c$, see the insets in Fig. \ref{fig:FitHom}. This may be explained as a consequence of the large fluctuations occurring in proximity of the transition: the subsystem at the boundary may act as a nucleation surface that facilitates the appearance of large scale paramagnetic domain that extends deep into the system instead of being confined near the boundary.

\paragraph{Non-Markovian Monte Carlo} -- We now turn to the study of a prototypical non-Markovian system.
\begin{figure}[!t]
    \centering
    \includegraphics[width=0.96\columnwidth]{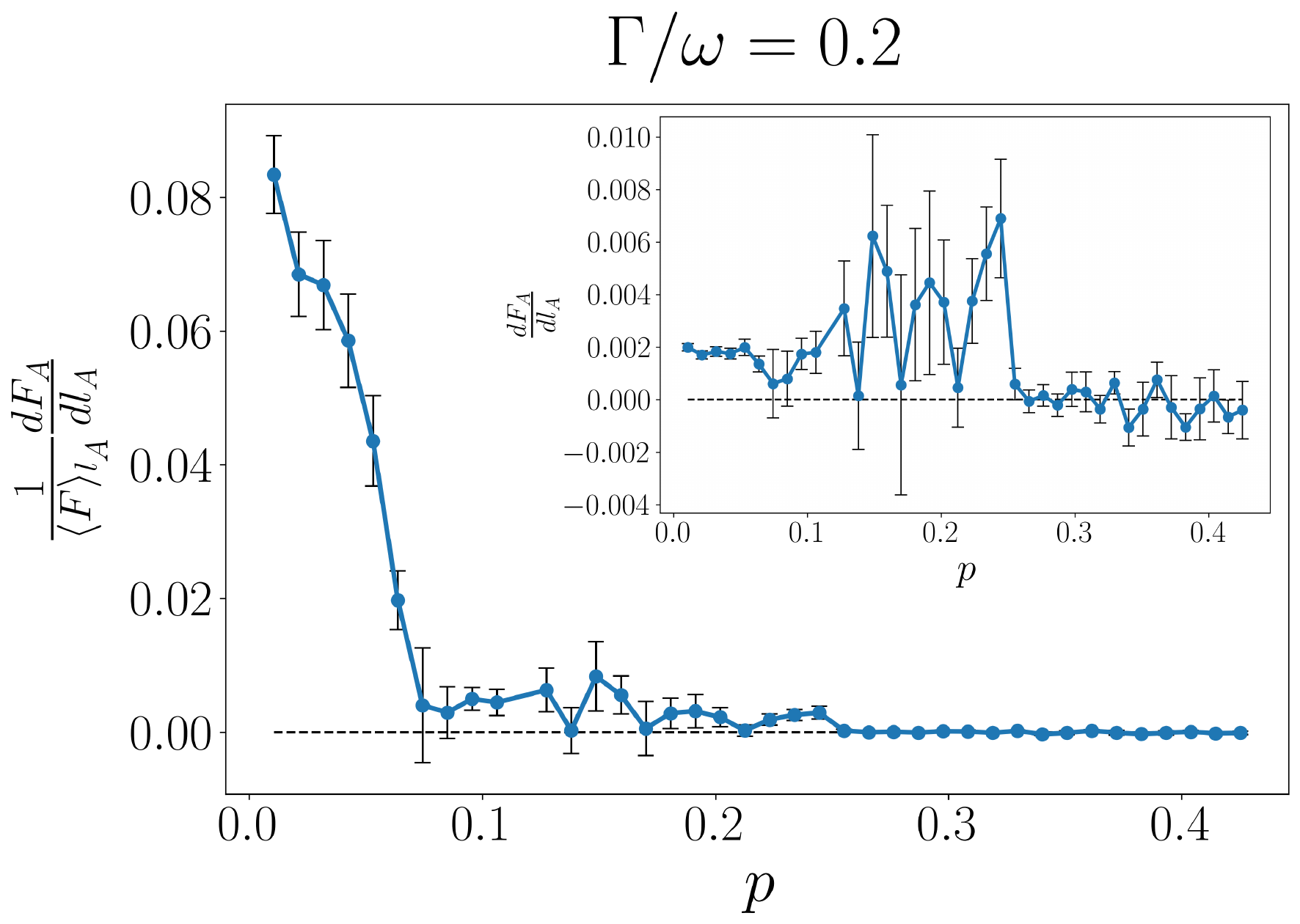}
    \caption{Behavior of the slope of the normalized energy $F_A/\langle F_A\rangle_{l_A}$ with respect to $l_A$ as function of $p$. The inset shows the slope of $F_A$ as function of $p$. The Monte Carlo calculations were performed for $\Gamma/\omega=0.2$, $L_x=40$, $L_y=50$ and $l_A=0,2,4,...,20$.}
    \label{fig:FitInh}
\end{figure}

We assume the decay rate $\Delta(t)$ to originate from a bath whose spectral density is described by a Lorentzian centered around $\omega$ and with bandwidth $\Gamma$. Within the time-convolutionless approximation \cite{Breuer:Petruccione,Piilo2009}, the decay rate appearing in the master equation \eqref{Eq:NMQMasterEquation} is
\begin{equation}\label{Eq:Deltat}
\Delta(t)=\Delta_0\left[\frac{\Gamma}{\omega}+e^{-\Gamma t}\left(\sin(\omega t)-\frac{\Gamma}{\omega}\cos(\omega t)\right)\right].
\end{equation}
This is a good approximation of a system of qudits, where each qudit level couples through its occupation number to a cavity mode with detuning $\omega$ and bandwidth $\Gamma$.

The rate decays to $\Delta_0\Gamma/\omega$ over a timescale $\sim1/\Gamma$ and has minima at $\omega t=3\pi/2+2\pi n$. The first (and lower) minimum is negative for $\Gamma/\omega<0.274$, meaning that $\Gamma$ sets the non-Markovianity of the dynamics.

The normalization constant $\Delta_0$ depends on the interaction strength. We choose each discrete time step in the $L_y$ direction to correspond to $\omega t=1/2$. We then set $\Delta_0$ so that the probability associated to the asymptotic value of $\Delta(t)$ is $p$
\begin{equation}\label{Eq:p_i}
p_i=p\left[1+e^{-\Gamma t_i}\left(\frac{\sin(\omega t_i)}{\Gamma/\omega}-\cos(\omega t_i)\right)\right].
\end{equation}
\begin{figure*}[!ht]
    \centering
    \includegraphics[width=0.96\textwidth]{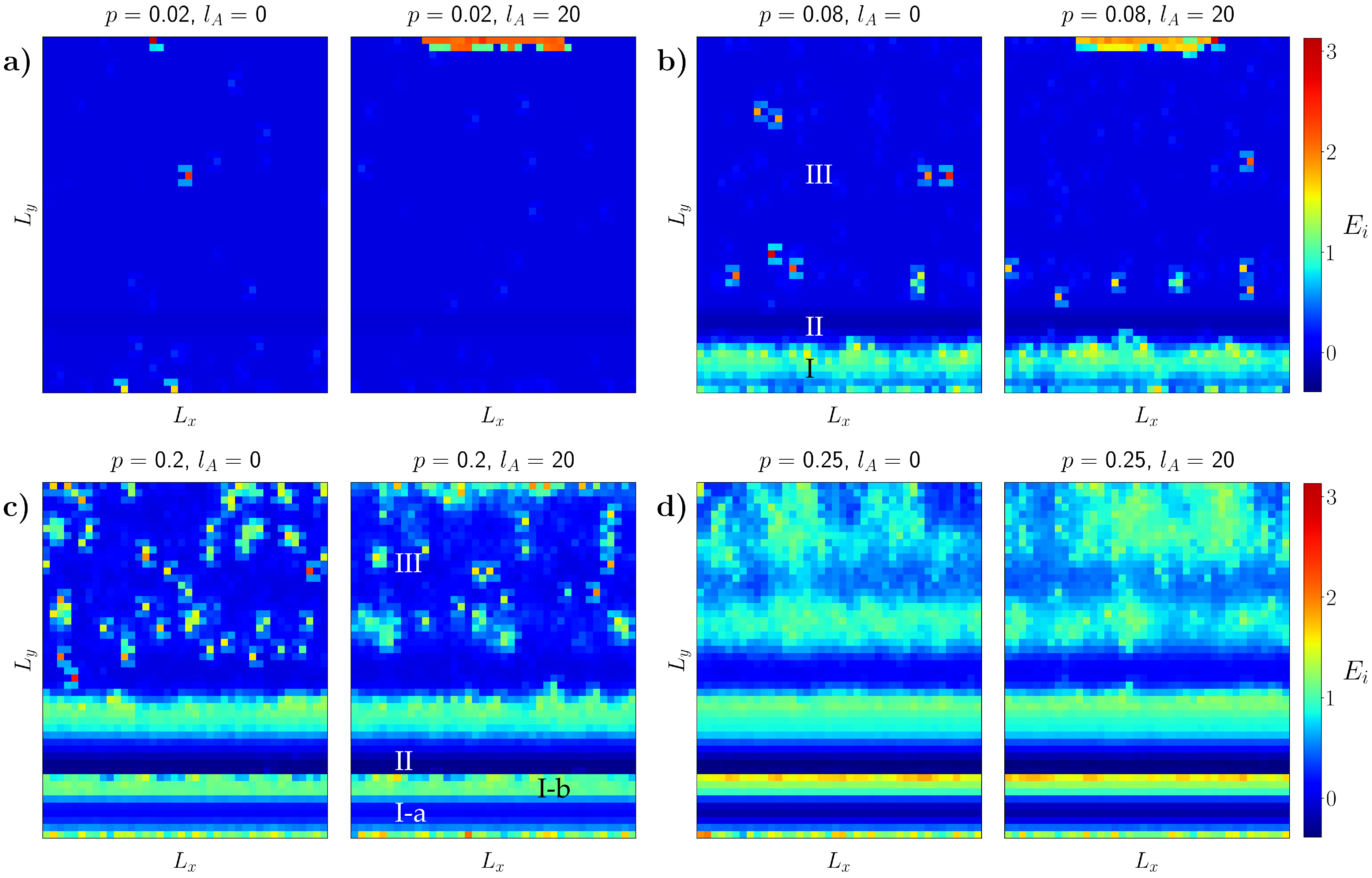}
    \caption{Colormap of the energy $E_i$ of each spin. Each panel shows $l_A=0$ or $l_A=20$ for (a) $p=0.1$, (b) $p=0.3$, (c) $p=0.4$, (d) $p=1$. Lower energy corresponds to aligned spins, i.e. ferromagnetic regions, while larger energy corresponds to paramagnetic spins. The labels $L_y$ in the vertical direction correspond to the time direction in the physical system. a) The system is completely ferromagnetic and the energy cost of having a boundary is clearly visible at the top of the right plot. b) The system is still ferromagnetic at later times (with the energy cost of the boundary still clearly visible) but small paramagnetic domains start to form at earlier times (when the peak value is $p_i\sim0.3\gtrsim p^c_{\textrm{hom}}$). c) The system is paramagnetic at early times since $p\sim0.4>p^c_{\textrm{hom}}$ contributing to the total energy, but the successive non-Markovian region is highly ferromagnetic and confines paramagnetism to early times. d) At late times $p_i\sim p^c_{\textrm{hom}}$, causing paramagnetic domains to appear after the non-Markovian region; the energy cost of the boundary is now small.}
    \label{fig:CMapInh}
\end{figure*}

We remark that the value of $p_i$ in Eq. \eqref{Eq:p_i} cannot be immediately mapped to the measurement probability $p$ of the Markovian case. They can only be compared in a sensible way at large times, where $p_i$ converges to a constant (and Markovian) measurement probability. We thus refrain from calling $p_i$ explicitly a probability. However, the earlier times behavior of the non-Markovian $p_i$ still affects the behavior of the system -- as we shall see in detail -- in a way that an analogy with the Markovian case cannot really be made.

We also note that depending on $p$ and $\Gamma/\omega$, $p_i$ can exceed one for certain times. This may seem weird, but is mathematically correct and corresponds to a coupling that favors a paramagnetic phase, since it gives a zero energy for aligned spins and an energy $\sim-\ln p_i<0$ for paramagnetic spins. This intuitively makes sense, since for very large $p_i$ -- i.e. very large decay rates -- the system tends to be paramagnetic rather than ferromagnetic.

We perform Monte Carlo simulations for $\Gamma/\omega=0.2$, which means the rate is negative for $3.77<\omega t<5.80$. Thus for a system with $L_y=50$ spins and $\omega t_i$ this means that the spins with $8\leq i_y\leq11$ have $p_i<0$, i.e. experience a non-Markovian coupling. The non-Markovian region is indicated with the label II in Fig. \ref{fig:CMapInh}.

Our results are reported in Fig. \ref{fig:FitInh} and \ref{fig:CMapInh}. We can immediately observe several similarities and some differences with the Markovian case.

The normalized slope $\frac1{\langle F_A\rangle_{l_A}}\frac{dF_A}{dl_A}$ drops from very large values to smaller values (but non zero) around $p=p_{c1}\approx0.08$; this corresponds to an increase of the fluctuations and of the slope $\frac{dF_A}{dl_A}$. For $p>p_{c2}\approx0.25$ both slopes decrease to zero, signalling an entanglement phase transition from volume to area law.

The strange phase between $p_{c1}$ and $p_{c2}$ corresponds to an emergence of paramagnetic domains at earlier times, before the non-Markovian region, indicated by the label I in Fig. \ref{fig:CMapInh}. Indeed, $p_{c1}$ corresponds to a peak value in the I-a region approximately equal to $p_i\sim0.4$, which is sufficient to turn paramagnetic the bottom region at early times, see Fig. \ref{fig:CMapInh}b. Consequently the energy of the system increases, which explains the drop in $\frac1{\langle F_A\rangle_{l_A}}\frac{dF_A}{dl_A}$, and the system is more susceptible to different boundary conditions, thus explaining the increase in $\frac{dF_A}{dl_A}$. However, the value of $p_i$ at later times (region III in Fig. \ref{fig:CMapInh}) is still too small to turn paramagnetic the top region, so that the system still exhibits a volume law behavior, as it is evident by the boundary energy cost in Fig. \ref{fig:CMapInh}c. 

For larger values $p\sim p_{c2}$ also the late times regions of the system start to turn paramagnetic, explaining the decrease of both slopes, see Fig. \ref{fig:FitInh} and \ref{fig:CMapInh}c. For $p$ larger than $p_{c2}$, the entire system turns paramagnetic, except for the non-Markovian region which is constrained to be ferromagnetic, see Fig. \ref{fig:CMapInh}d.

We also notice that for $p\gtrsim0.15$ the region I at earlier times exhibits two energy subregions I-a and I-b, see Fig. \ref{fig:CMapInh}c. These subregions are both paramagnetic, but I-a has a lower energy because $p_i$ exhibits its peak in I-a; this large probability lowers the energy of the paramagnetic phase. In I-b, $p_i$ decreases and eventually vanishes before becoming negative in region II; thus the energy of the paramagnetic phase increases as $p_i$ decreases, explaining the different energy behavior inside of region I.

The phase between $p_{c1}$ and $p_{c2}$ is still volume law despite exhibiting large energy fluctuations. The width of this region is likely size dependent, since evolving the system for longer times would suppress the influence of the paramagnetic region (I) and of the non-Markovian region (II) at early times on the late times (III) region.

Indeed, the transition from volume law to area law is mostly determined by the late times values of $p_i$ and only occurs at $p_{c2}\approx0.25$, similarly to the Markovian transition. This confirms the intuition that the late times non-Markovian dynamics, when the rate is always positive $\Delta(t)>0$, is equivalent to a Markovian dynamics. 

An interesting result is that the volume law phase still survives even when the peak value of $p_i$ becomes \textit{significantly} larger than the Markovian critical probability. This occurs because while the peak $p_i$ is large enough to turn paramagnetic the early times region (I), the successive non-Markovian region (II) is always ferromagnetic and shields the rest of the evolution from the effects of this large peak value.

We conclude that, while non-Markovianity does not affect the volume law phase at late times, when most of the dynamics has become Markovian, it stabilizes the volume law phase at early times and protects it from regions of strong measurements, provided they occur before the non-Markovian region.

We remark that while the numerical results obtained with our Monte Carlo simulations display  fluctuations, especially near the transition, they still provide a qualitative (and somewhat quantitative) picture of the Markovian and non-Markovian transition. Precision can be improved by increasing the system size and the number of sampling steps, but this has a somehow high computational cost, particularly since we have to utilize boundary conditions that make Monte Carlo simulations slower compared to a system with periodic boundary conditions. 

\section{Conclusions}\label{Sec:Conclusions}

We have introduced a theoretical framework to unravel the non-Markovian dynamics of quantum many-body systems in terms of quantum trajectories interspersed by quantum jumps. Our technique relies on two methodological innovations: a formulation of many-body quantum jumps applicable to certain classes of non-Markovian dynamics, and a diagrammatic expansion to map the resulting evolution into amenable equations of motion.

Unlike in the Markov case, non-Markovian many-body trajectories are not independent from each other, a direct consequence of the fact that the bath retains finite memory due to non-trivial spectral functions. This features makes averaging the system dynamics from trajectories practically intractable at the computational level.

The key feature of our framework is that it allows to investigate measurement induced phase transitions in the presence of information back-flow - a situation relevant to any system where measurements are realized via coupling to a non-trivial bath. This can be done analytically because, under mild assumptions (i.e. sufficiently large sizes and evolution times), it is possible to write down closed-form equations governing the conditional probability of each trajectory using diagrammatic methods. Remarkably, these equations share the same functional form of the Dyson equation, and can be manipulated so that the probability of a generic outcome trajectory is given in terms of the time evolution of the expectation value of local observables. Within our framework, this result shows how the - highly-non linear -  effect of non-Markovianity on many-body systems can be cast as a ``dressing'' over Markovian trajectories, very much like interactions do for single particle wave-functions in electronic systems. 

This Dyson equation-like description enables the study of entanglement and measurement induced transitions in the presence of information inflow from the bath back to the system. For the case of one-dimensional Haar circuits, we formulate a classical statistical mechanics model of the system dynamics. The key difference with the memory-less case is that couplings are now time-dependent, and that there are large regions of space time where magnetic fluctuations are suppressed: these are actually the non-Markovian regions, and their coupling profiles reflects the fact that reverse jumps are included in the statistical mechanics model implicitly via the dressed probability distributions. We study the properties of entanglement via numerical simulations of the Potts model describing the $N=3$ replica space. We point out that our approach is also applicable to the case of evolution with Clifford gates, where statistical mechanics mappings have been recently proposed \cite{Li2021:CliffordModel}.

Overall, our results demonstrate a previously unproved inherent robustness of measurement-induced transition to information backflow. Combined with the, by now established, fact that such transitions can occur for various kinds of coherent dynamics and Kraus operators, this suggests that measurement induced transitions might indeed take place in a variety of settings, including systems where the effects of information back-flow are often non-negligible. 

It is worth pointing out a few possible questions that our work raises. In terms of relevance to experiments, it would be important to combine our approach with an inherently open system description of the system (i.e. noise in addition to measurement), that, for the Markovian case, has been recently addressed in Ref.~\cite{weinstein2022measurement}. Moreover, our methods may find application in studying measurement induced transitions in solid state systems, where memory effects are important; for example, the $1/f$ noise is non-Markovian and is the most common type of noise in quantum devices based on solid state platforms \cite{Paladino2013:1fnoise}.

We also remark that the trajectories we consider in our work can still be realized through a (possibly very complicated) experimental setup, for example through a combination of quantum simulation of the system and a classical memory that stores information about the occurrence of the normal jumps~\cite{noel2022measurement,Koh2022:IBM}. This classical memory is essential to supply the memory effects of the non-Markovian evolution and allows to provide memory feedback in a controlled way, but it also requires exponentially large resources. Another way of realizing non-Markovian trajectories in a physical system, is by coupling the system of interest to an auxiliary bath with a non trivial dynamics and subjected to Markovian measurements, which results in an effective non-Markovian dynamics of the system. The evolution of system+bath is described by conventional quantum trajectories, and in the case where the measurements act globally on the bath, the system evolves along a pure state trajectory for which the formalism of our paper is directly applicable. Indeed, we have illustrated the above idea within the context of coupled free-fermion chains in a recent work \cite{Tsitsishvili2023}, where we also confirmed numerically a key qualitative prediction of the diagrammatic approach -- that is the ’stability’ of the measurement-induced transition.

Within the context of measurement induced transitions, another question is about the connection between error correction schemes and measurement protocols. Given the fact that memory effects can in principle be precisely quantified in experiments by performing spectroscopy of the bath, it would be interesting to see whether that information can be utilized to improve error correction, or at least, if the presence of a measurement induced transition can at least provide some intrinsic robustness of decoding methods with respect to measurement errors (that can also be seen as a non-Markovian effect in some cases) \cite{Niroula2023:error}. 

Another open question is the formulation of practical numerical procedures to address MIPT in the presence of memory. Here, it might be possible to adapt some methods that have found success in few body systems, at least for a qualitative understanding. Beyond such applications, it would be interesting to see whether our diagrammatic method can provide insights on other many-body phenomena in non-Markovian systems, as well as on recently developed computational techniques to tackle them~\cite{Rivas2014:NM_measuresReview}.

\section*{Acknowledgements}

We thank M. Buchhold, H. P. B\"uchler, J. Piilo, M. Schir\'o, S. Trebst, V. Vitale, M. Fabrizio and P. Zoller for insightful discussions. The work of G.C., M.D. and M.T. is partly supported by the ERC under grant number 758329 (AGEnTh) and by the MIUR Programme FARE (MEPH). M. T. thanks the Simons Foundation for supporting his Ph.D. studies through Award 284558FY19 to the ICTP. D.P. acknowledges support from the National Research Foundation, Singapore under its QEP2.0 programme (NRF2021-QEP2-02-P03). G.C. acknowledges financial support from by ICSC–Centro Nazionale di Ricerca in High-Performance Computing, Big Data and Quantum Computing. The work of R.F. is co-funded by the European Union (ERC, RAVE, 101053159). Views and opinions expressed are however those of the authors only and do not necessarily reflect those of the European Union or the European Research Council. Neither the European Union nor the granting authority can be held responsible for them. We also acknowledge financial support from PNRR MUR project PE0000023-NQSTI.

\bibliographystyle{apsrev4-1}
\bibliography{biblio.bib}

\end{document}